\documentclass[11pt]{article}
\usepackage[a4paper, margin=1.05in]{geometry}
\usepackage{url}
\usepackage{breqn}
\usepackage{algorithm}
\usepackage{algpseudocode}
\usepackage{amsthm}
\usepackage[round]{natbib}
\usepackage{authblk}
\usepackage{xcolor}
\usepackage{hyperref}

\theoremstyle{definition}

\begin{document}

% Title of paper
\title{Multi-omics network reconstruction with\\collaborative graphical lasso}

% List of authors, with corresponding author marked by asterisk
\author[1,2]{Alessio Albanese}
\author[2]{Wouter Kohlen}
\author[1]{Pariya Behrouzi*}
\affil[1]{Mathematical and Statistical Methods group - Biometris,
Wageningen University and Research}
\affil[2]{Cell and Developmental Biology,
Wageningen University and Research}
\date{}

\renewcommand\Affilfont{\itshape\small}
\renewcommand*{\thefootnote}{\fnsymbol{footnote}}
%\author{ALESSIO ALBANESE\\[4pt]
%% Author addresses
%\textit{Mathematical and Statistical Methods (Biometris) and Cellular and Developmental Biology,
%Wageningen University and Research,
%% Droevendaalsesteeg 1,
%6708 PB,
%Wageningen,
%The Netherlands}
%\\
%% E-mail address for correspondence
%{alessio.albanese@wur.nl}\\[2pt]
%WOUTER KOHLEN\\[4pt]
%% Author addresses
%\textit{Cellular and Developmental Biology,
%Wageningen University and Research,
%% Droevendaalsesteeg 1,
%6708 PB,
%Wageningen,
%The Netherlands}\\[2pt]
%PARIYA BEHROUZI$^\ast$\\[4pt]
%% Author addresses
%\textit{Mathematical and Statistical Methods (Biometris),
%Wageningen University and Research,
%% Droevendaalsesteeg 1,
%6708 PB,
%Wageningen,
%The Netherlands}}
%\author{PARIYA BEHROUZI\\[4pt]
%% Author addresses
%\textit{Mathematical and Statistical Methods (Biometris),
%Wageningen University and Research,
%Droevendaalsesteeg 1,
%6708 PB,
%Wageningen,
%The Netherlands}
%}

% Running headers of paper:
%\markboth%
% First field is the short list of authors
%{A. Albanese, W. Kohlen and P. Behrouzi}
% Second field is the short title of the paper
%{Collaborative graphical lasso}

\maketitle

% Add a footnote for the corresponding author if one has been
% identified in the author list
\footnotetext{\textsuperscript{*}To whom correspondence should be addressed. Email: pariya.behrouzi@wur.nl}

\begin{abstract}
{\textbf{Motivation:} In recent years, the availability of multi-omics data has increased substantially. Multi-omics data integration methods mainly aim to leverage different molecular layers to gain a complete molecular description of biological processes. An attractive integration approach is the reconstruction of multi-omics networks. However, the development of effective multi-omics network reconstruction strategies lags behind.\\
\textbf{Results:} In this study, we introduce \textit{collaborative graphical lasso}, a novel approach that extends \textit{graphical lasso} by incorporating collaboration between omics layers, thereby improving multi-omics data integration and enhancing network inference.
Our method leverages a collaborative penalty term, which harmonizes the contribution of the omics layers to the reconstruction of the network structure. This promotes a cohesive integration of information across modalities, and it is introduced alongside a dual regularization scheme that separately controls sparsity \textit{within} and \textit{between} layers. To address the challenge of model selection in this framework, we propose \textit{XStARS}, a stability-based criterion for multi-dimensional hyperparameter tuning.
We assess the performance of \textit{collaborative graphical lasso} and the corresponding model selection procedure through simulations, and we apply them to publicly available multi-omics data. This application demonstrated \textit{collaborative graphical lasso} 
recovers established biological interactions while suggesting novel, biologically coherent connections.\\
\textbf{Availability and implementation:} We implemented \textit{collaborative graphical lasso} as an R package, available on CRAN as \textit{coglasso}. The results of the manuscript can be reproduced running the code available at \url{https://github.com/DrQuestion/coglasso_reproducible_code}.} \\[4pt]
\end{abstract}

\section{Introduction}
\label{sec1}

A successful multi-omics data integration strategy would provide a holistic molecular explanation of any biological phenomenon of interest. Therefore, multi-omics data integration can be considered a paramount topic of computational biology. Two well known strategies for multi-omics data integration include Multi Omics Factor Analysis (\textit{MOFA}, \cite{argelaguet_mofa_2020}) and Data Integration Analysis for Biomarker discovery using Latent components (\textit{DIABLO}, \cite{singh_diablo_2019}). \textit{MOFA} is an generalization of Bayesian factor analysis to allow the study of the multiple data layers that are typical of the multi-omics setting. \textit{DIABLO}, instead, extends sparse canonical correlation analysis to a supervised framework, allowing to identify the multi-omics variables that best predict the phenotype under study.

Another attractive strategy to integrate multi-omics data is to estimate networks of interactions from them, connecting the molecular units under investigation. These estimated interactions hint directly at the regulatory mechanisms that underlie the biological phenomenon during which multi-omics data were recorded. A preferred statistical framework to reconstruct these networks is Gaussian Graphical Models (GGMs),
which offer a powerful framework for inferring conditional dependencies among molecular entities. The widely used \textit{graphical lasso} (\textit{glasso}) method efficiently estimates sparse GGMs by introducing an $L_1$ regularization penalty, ensuring sparsity in the precision matrix \citep{friedman_sparse_2008}. However, \textit{glasso} is designed for single-layer applications and does not account for the distinct yet interdependent nature of multi-omics data. A first effort to adapt the GGM framework to multi-omics data integration is Determiming Regulatory Associations using Graphical models on multi-Omics Networks (\textit{DRAGON}, \cite{shutta_dragon_2023}). \textit{DRAGON} extends the $L_1$ regularization of penalized GGM estimation with the introduction of separate penalty parameters for the different omics layers.

While the field of multi-omics GGMs estimation is still in its infancy, researchers have already developed several prediction techniques based on linear regression to integrate multi-omics data. Some of these techniques have introduced concepts that could strongly benefit current GGM estimation methods. 
For example, \citet{gross_collaborative_2015} introduced \textit{collaborative regression}, a strategy that handles the multi-layer nature of multi-omics data in a remarkable way. This technique encourages ``collaboration'' between two omics layers by penalizing the difference between the linear predictors obtained from each of the two layers. In this way, the two layers comparably contribute to the prediction. Despite this original and elegant way to deal with the multi-layer nature of multi-omics data, the concept of collaboration has not been implemented in the field of GGM estimation.

To bridge this gap, we propose \textit{collaborative graphical lasso} (\textit{coglasso}), which extends \textit{glasso} by introducing a collaborative penalty term. This term simultaneously balances the contributions of the omics layers to the GGM estimation and encourages shared information among them.
Along with collaboration, we introduce a system of two independent penalty parameters, a \textit{within} parameter ($\lambda_w$) and a \textit{between} parameter, ($\lambda_b$). These two can vary independently of each other in a range of values, allowing to penalize differently connections within the same class and between different classes of variables. To select the best combination of these hyperparameters we designed \textit{XStARS}, a stability-based criterion that extends \textit{StARS} \citep{liu_stability_2010} to select the best \textit{coglasso} network. The combination of collaboration and mixed penalty parameters enables \textit{coglasso} to reconstruct a single integrated network from two omics layers, each containing a different set and class of variables that are encouraged to borrow information from each other.

Here, we assess the performance of our method in extensive simulation studies comparing it with the \textit{glasso} method. Further, we show usefulness of \textit{coglasso} with an application on a multi-omics study of sleep deprivation. Our method is available on CRAN as the R package \textit{coglasso}, while the reproducible code to replicate the simulations and the analysis of the multi-omics dataset of sleep deprivation is available at \url{https://github.com/DrQuestion/coglasso_reproducible_code}.

\section{Materials and Methods}
\label{sec2}

\textit{Collaborative graphical lasso }(\textit{coglasso}) is a novel algorithm to estimate gaussian graphical models (GGMs) from data with a multi-layer nature, \textit{e.g.} multi-omics data. To develop this new algorithm, we extended the \textit{glasso} algorithm proposed by \citet{friedman_sparse_2008}, which is exclusively designed for single-layer applications. The resulting algorithm can estimate GGMs by allowing variables that belong to different omics layers to borrow information from each other. The approach, designed to leverage these different classes of variables, is inspired by \textit{collaborative regression} introduced by \citet{gross_collaborative_2015}. Similarly to how \textit{collaborative regression} integrates multiple omics layers by modifying the objective function of the basic linear regression, \textit{coglasso} achieves a similar level of integration by modifying the objective function at the basis of the \textit{glasso} algorithm. 

\subsection{Gaussian Graphical Models}\label{subsec:ggm_glasso}

Gaussian Graphical Models (GGMs) are a powerful framework to represent variables distributed according to a multivariate normal distribution and their conditional dependence relationships. They can be represented by graphs. A graph $G=(V, E)$ is defined by a set of nodes $V$ and a set of edges $E$ that represent connections between pairs of nodes. In the GGM framework, nodes are the variables of the model, while edges, the connections between such variables, are used to represent the conditional dependence relationships between them. This means that two variables are connected if they are conditionally dependent, while they are not connected if they are conditionally independent. Assume, for example, that we are interested in estimating a GGM with $p$ variables. We assume that these variables follow a multivariate normal distribution with mean $\mu$ and variance-covariance matrix $\boldsymbol{\Sigma}$. We call the inverse of a variance-covariance matrix $\boldsymbol{\Sigma}$ a precision matrix $\boldsymbol{\Theta}=\boldsymbol{\Sigma}^{-1}$. The entries of a precision matrix represent conditional dependencies between variables. The conditional dependence between two variables can be interpreted as what remains of the dependence between the two after removing the role that all the other variables of the system have on that dependence. In particular, if the entry $\theta_{ij}$ is zero it means that the variables $i$ and $j$ are conditionally independent, given all the other variables in the system. Therefore, since in the GGM framework connections represent conditional dependence relationships, the set of connections of a GGMs is purely determined by non-zeros in the precision matrix. These connections can be further represented in the adjacency matrix $\boldsymbol{A}$, where all non-zero elements (the edges of the network) are encoded as $1$s and the remaining ones as $0$s. Another common way to represent the entries of a precision matrix is in the form of partial correlations, computed for the entry $(i,j)$ as $\frac{\theta_{ij}}{\sqrt{\theta_{ii}\theta_{jj}}}$.

A well-established algorithm to estimate GGMs from a single omics layer is \textit{glasso} \citep{friedman_sparse_2008}. This algorithm performs an iterative coordinate descent optimizing procedure to find $\boldsymbol{W}$, the penalized estimate of the variance-covariance matrix $\boldsymbol{\Sigma}$, and its inverse $\widehat{\boldsymbol{\Theta}}$. \textit{Glasso}’s strategy allows positive definiteness, hence invertibility, of the matrix $\boldsymbol{W}$ resulting from the estimation. Moreover, \textit{glasso} guarantees the inverse matrix $\widehat{\boldsymbol{\Theta}}=\boldsymbol{W}^{-1}$ to be sparse, meaning that the final GGM determined by $\widehat{\boldsymbol{\Theta}}$ is not too dense with connections. \textit{Glasso} achieves these positive definiteness and sparsity by fragmenting the estimation of $\boldsymbol{W}$ into several \textit{lasso} regressions that optimize and estimate separately each row and column of the matrix. Every \textit{lasso} regression solves Equation \eqref{eqn:obj_glasso}.
\begin{equation}\label{eqn:obj_glasso}
\boldsymbol{\widehat{\beta}_i}=\arg \min_{\boldsymbol{\beta_i}} \left \{ \frac{1}{2}\left \| (\boldsymbol{W}_{\setminus i\setminus i})^{-1/2}\boldsymbol{S}_i-(\boldsymbol{W}_{\setminus i\setminus i})^{1/2}\boldsymbol{\beta_i}\right \|^2 + \lambda \left \| \boldsymbol{\beta_i} \right \|_1\right \}
\end{equation}
Here $\boldsymbol{S}$ is the empirical variance-covariance matrix estimated from the data and $\lambda$ is the penalty parameter. $\| \cdot \|_1$ and $\| \cdot \|^2$ represent, respectively, the $L_1$- and the $L_2$-norm. $\boldsymbol{S}_i$ is the $i$-th column (and row) of $\boldsymbol{S}$ without the diagonal element, while $\boldsymbol{W}_{\setminus i \setminus i}$ is the submatrix of $\boldsymbol{W}$ without the $i$-th row and column. The vector $\boldsymbol{\widehat{\beta}_i}$ obtained solving equation \eqref{eqn:obj_glasso} represents a proxy for the connections to the $i$-th node. When the $j$-th entry of the vector is zero, then the nodes $i$ and $j$ are not connected. Once obtained, $\boldsymbol{\widehat{\beta}_i}$ is stored as the $i$-th column of the matrix $\widehat{\boldsymbol{B}}$ and used to update the $i$-th row (and column) of $\boldsymbol{W}$, $\boldsymbol{W}_i$, as $\boldsymbol{W}_i= \boldsymbol{W}_{\setminus i \setminus i} \boldsymbol{\beta_i}$. $\boldsymbol{W}_{ii}$, the $i$-th diagonal element of the matrix $\boldsymbol{W}$, is initiated to $\boldsymbol{W}_{ii} = \boldsymbol{S}_{ii} +\lambda$ at the beginning of the algorithm. The algorithm cycles through $i=1,2,\ldots,p,1,2,\ldots,p,\ldots$ until convergence, updating $\boldsymbol{\beta_i}$ and $\boldsymbol{W}_i$ in each iteration.

For any $i$, $\boldsymbol{\beta_i}$ is estimated element by element through a coordinate descent procedure as described in \citet{friedman_pathwise_2007}. This procedure iteratively updates the $j$-th coordinate of the vector $\boldsymbol{\beta_i}$, $(\boldsymbol{\widehat{\beta}_i})_j$, which is the main proxy for the conditional dependence between node $i$ and node $j$ of the network. We report the update rule to update $(\boldsymbol{\widehat{\beta}_i})_j$ in Algorithm 1 of the Supplementary Data.

\subsection{\textup{Collaborative graphical lasso}}\label{sec3}

\textit{Collaborative graphical lasso} (\textit{coglasso}) combines the strengths of \textit{collaborative regression} with the well-established method for GGM estimation \textit{glasso}. This integration is possible by incorporating the former’s ability to leverage multiple omics layers within the latter's framework. Let us have two omics layers, $\boldsymbol{X}$ and $\boldsymbol{Z}$ (\textit{e.g.}, a transcriptomic and a metabolomic layer), respectively containing two distinct kinds of variables of size $p_x$ and $p_z$ measured over the same set of samples. Let $p= p_x + p_z$ be the total number of variables. 
Like \textit{glasso}, \textit{coglasso} breaks the estimation of the variance-covariance matrix $\boldsymbol{W}$ into multiple lasso regressions, each one leading to the update of a row and column of the matrix $\boldsymbol{W}$. These multiple lasso regressions guarantee that the resulting $\boldsymbol{W}$ is invertible and that its inverse matrix is sparse. 
However, a key limitation of applying \textit{glasso} in multi-layer scenarios is its inability to account for distinct classes of variables, which commonly arise in multi-omics studies. Using \textit{glasso} would require merging $\boldsymbol{X}$ and $\boldsymbol{Z}$ as though they measured the same type of variable, potentially losing critical layer-specific information.
In contrast, \textit{coglasso} introduces an innovative approach to account for the unique contributions of each omics layer when estimating a GGM. This is achieved through a fundamental modification of \textit{glasso}’s objective function. First, \textit{coglasso} incorporates a collaborative term that explicitly separates and leverages the contributions of $\boldsymbol{X}$ and $\boldsymbol{Z}$. Second, it applies a mixed system of  $L_1$ penalties:
one penalizing the interactions \textit{within} the variables of the same omics layer and one for the interaction \textit{between} different omics layers. After these elements are incorporated, \textit{coglasso}’s objective function of the $i$-th regression, estimating the proxy for the connections to the $i$-th variable $\boldsymbol{\widehat{\beta}_i}$, and hence the $i$-th row and column of $\boldsymbol{W}$, takes the form

\begin{align}\label{eqn:obj_coglasso}
    \boldsymbol{\widehat{\beta}_i}= \arg \min_{\boldsymbol{\beta_i}}\left \{  \frac{1}{2}\left \| (\boldsymbol{W}_{\setminus i\setminus i})^{-1/2}\boldsymbol{S}_i-(\boldsymbol{W}_{\setminus i\setminus i})^{1/2}\boldsymbol{\beta_i}\right \|^2\right. \nonumber
    \\  + \frac{c}{2}\left \| (\boldsymbol{W}_{\setminus i\setminus i})^{1/2}_X(\boldsymbol{\beta_i})_X- (\boldsymbol{W}_{\setminus i\setminus i})^{1/2}_Z(\boldsymbol{\beta_i})_Z\right \|^2 \nonumber
    \\  \left.+ \left \| \boldsymbol{\Lambda}_i \odot \boldsymbol{\beta}_i \right \|_1\right \}
\end{align}

In Equation \eqref{eqn:obj_coglasso}, beside the typical \textit{glasso} term, we introduce a collaborative term. In this term, we partition the $\boldsymbol{W}_{\setminus i\setminus i}$ sub-matrix and the vector of coefficients $\boldsymbol{\beta_i}$ from Equation \eqref{eqn:obj_glasso} into two main components, one for each omics layer. We define the partitions as follows: 
\begin{equation*}
\boldsymbol{W}_{\setminus i\setminus i} = \begin{pmatrix}
(\boldsymbol{W}_{\setminus i\setminus i})_X, & (\boldsymbol{W}_{\setminus i\setminus i})_Z
\end{pmatrix} \quad \textrm{and} \quad
\boldsymbol{\beta_i} = \begin{pmatrix}
(\boldsymbol{\beta_i})_X \\
(\boldsymbol{\beta_i})_Z
\end{pmatrix}.
\end{equation*}

Here, every $j$-th column of $\boldsymbol{W}_{\setminus i\setminus i}$ is directly related to the contribution of the single $j$-th variable to the network structure. \textit{Coglasso} partitions these contributions into two different groups: $(\boldsymbol{W}_{\setminus i\setminus i})_X$ and $(\boldsymbol{W}_{\setminus i\setminus i})_Z$. These contain the contributions of the variables of omics layer $\boldsymbol{X}$ and of the omics layer $\boldsymbol{Z}$, respectively. For a detailed description of the dimensionalities of these partitions, please refer to the Supplementary Data. Analogously to the $\boldsymbol{W}_{\setminus i\setminus i}$ matrix, \textit{coglasso} partitions the coefficients contained in $\boldsymbol{\beta_i}$. Each $j$-th coefficient represents a proxy for the connection between the $i$-th and the $j$-th variable, and the weight given to the contribution to the network structure of the $j$-th variable when solving Equation \eqref{eqn:obj_coglasso} for the $i$-th variable. \textit{Coglasso} partitions also these coefficients into two groups, one per omics layer. As a consequence, in Equation \eqref{eqn:obj_coglasso}
$(\boldsymbol{W}_{\setminus i\setminus i})^{1/2}\boldsymbol{\beta_i}$, representing the weighted contribution of every variable to the network structure, is also partitioned in the two sub-elements whose distance is minimized by the collaborative term. 

Hence, the Equation \eqref{eqn:obj_coglasso} is made of three components. The first term coincides with the main term of \textit{glasso}'s objective function, and it is responsible for the estimation of the network structure. The second term, the collaborative term, identifies the separate contributions of omics layer $\boldsymbol{X}$ and of the omics layer $\boldsymbol{Z}$ to the network estimation, encouraging them to borrow information from each other. The last term in Equation \eqref{eqn:obj_coglasso} is the penalty term. Here, the $\odot$ symbolizes the Hadamard product, and $\boldsymbol{\Lambda}$ is the matrix of penalty parameters that allows to differently penalize \textit{within} layer and \textit{between} layer interactions. It is a $p \times p$ matrix with two square diagonal blocks of dimensions $p_x\times p_x$ and $p_z \times p_z$, respectively, and two off-diagonal blocks of dimensions $p_x \times p_z$ and $p_z\times p_x$. The diagonal blocks contain the penalty parameter regulating the within-layer interactions, $\lambda_w$, while the off-diagonal blocks contain the parameter penalizing the between-layer interactions, $\lambda_b$. In practice, the former controls the density of connections between nodes belonging to the same omics layers, while the latter controls the density of the connections between nodes belonging to different omics layers.

To further illustrate the role of the collaborative term, let us consider the estimation of a conditional independence graph as a redistribution of the information contained in every variable $i$ among all other variables of the system. Traditional \textit{graphical lasso} performs this redistribution iteratively, assigning the information in variable $i$ only to the most explanatory variables due to the sparsity induced by the $L_1$-penalty. However, this approach ignores the distinct sources of variables in a multi-omics setting. 
The collaborative term in \textit{coglasso} ensures that the redistribution of the information contained in the variable $i$ is harmonized across the different omics layers, preventing omics layer-specific biases from dominating the network inference process and promoting a more balanced integration of multi-omics data. The hyperparameter regulating the intensity of collaboration is $c$, the collaboration value. It is straightforward to notice that for $c=0$, \textit{coglasso} removes omics layer-harmonization, but still applies the mixed penalty system. 
By incorporating these elements, \textit{coglasso} provides a structured yet flexible approach for multi-omics network inference that is interwoven with the share of information among the omics layers.

We solve Equation \eqref{eqn:obj_coglasso} iteratively to obtain $\boldsymbol{\widehat{\beta}_i}$ for every variable $i=1,2,\ldots,p,1,2,\ldots,p,\ldots$ until convergence. At every iteration this vector is used to update $\boldsymbol{W}$ as described in Section \ref{subsec:ggm_glasso} ($\boldsymbol{W}_i= \boldsymbol{W}_{\setminus i\setminus i} \boldsymbol{\beta_i}$), then it is stored as the $i$-th column of a matrix $\boldsymbol{\widehat{B}}$. 
To estimate the $j$-th element of $\boldsymbol{\beta_i}$, which represents a proxy for the connection between the $i$-th and the $j$-th variable, \textit{coglasso} adopts a coordinate descent procedure. We derived the coordinate update rule from Equation \eqref{eqn:obj_coglasso}, and we present it in Algorithm 2 of the Supplementary Data. 

\subsection{Stability selection for \textup{collaborative graphical lasso}}

\textit{Coglasso} requires the setting of three hyperparameters: $\lambda_w$, $\lambda_b$ and $c$. The optimal choice of hyperparameters can be a challenging task that is usually tackled with model selection procedures. An attractive model selection procedure developed to choose the best $\lambda$ for \textit{glasso} is \textit{StARS} \citep{liu_stability_2010}. Here is how StARS works in principle: it first builds a network for every value of $\lambda$ in a descending array of values, gradually moving from sparser networks to denser networks. For every $\lambda$ value, \textit{StARS} computes the average edge stability of the associated network under repeated subsamplings of the original dataset. Very sparse, near-empty networks will be associated to a high stability, as a very few edges will be allowed. For decreasing $\lambda$ values, and hence for less sparse networks, the stability will gradually decrease, as more edges will be allowed to vary. This will be until the obtained networks are so dense that the same edges will keep appearing across subsamplings, leading to a renewed edge stability. As \textit{StARS} favors sparse networks, it eventually will select the highest $\lambda$ value in the descending order for which the stability is still above a given threshold. However, to select the optimal \textit{coglasso} network with a comparable approach, \textit{StARS} needs to be adapted to explore a hyperparameter space that spans not one, but three dimensions. All these three dimensions, those of $\lambda_w$, $\lambda_b$ and $c$, influence, in varying measure and direction, the sparsity of a network, hence its stability. In particular, as $\lambda_w$, $\lambda_b$ increase, the network become progressively sparser. As $c$ increases, instead, the network gains more connections. We designed an alternative version of \textit{StARS} able to explore the three-dimensional hyperparameter space of \textit{coglasso}, naming it \textit{eXtended StARS} (\textit{XStARS}), and we describe it in Algorithm 3 of the Supplementary Data.

\subsection{Implementation} 
We implemented \textit{coglasso} algorithm in C++, and we developed an R package as an interface to it, naming it \textit{coglasso}. The package also implements the selection algorithm \textit{XStARS} described above. \textit{Coglasso} is currently distributed through CRAN, and, for the C++ implementation, we were inspired by \citet{zhao_huge_2012}.

\section{Results}\label{sec4}
\subsection{Simulations}
We conducted simulations to assess \textit{coglasso}’s ability to accurately recover underlying network structures. In this evaluation, \textit{coglasso} was compared with the well-established original \textit{glasso} algorithm from \citet{friedman_sparse_2008}.
We evaluated the methods across three scenarios, each presenting a network reconstruction challenge of increasing structural complexity. In each scenario, we first generated a GGM with a complex covariance structure designed to resemble two omics layers. We then used the GGM to simulate multi-omics datasets, which served as input for both methods to assess their performance.

For the generation of a multi-omics-like GGM for each scenario, we started from building the precision matrix, $\boldsymbol{\Theta}$. In each scenario, we split the generation of $\boldsymbol{\Theta}$ in three sub-tasks: the first two to generate the two diagonal blocks of the matrix $\boldsymbol{\Theta}_{XX}$ and $\boldsymbol{\Theta}_{ZZ}$, and the last one to generate the off-diagonal block $\boldsymbol{\Theta}_{XZ}$, that would finally be transposed to generate $\boldsymbol{\Theta}_{ZX} = \boldsymbol{\Theta}_{XZ}^T$ and to impose symmetry over the matrix.

We started by generating the two diagonal blocks of the precision matrix, one for each simulated omics layer, by generating their \textit{within}-layer networks. These two networks of the two omics layers were always of two different sizes, and we simulated them via the R package \textit{huge} as cluster networks. The first and largest layer was always modeled as a three-clusters network, with its size increasing across scenarios. In scenario $1$, the first layer contained $p_X=40$ nodes, where each node had a probability of $\frac{1}{4}$ of connecting to another node withing the same cluster. To avoid the three clusters being completely disjoint, we added $7$ random connections between them. In scenario $2$, the size $p_X$ increased to $80$ nodes, and the probability of within-cluster connection reduced to $\frac{1}{6}$. Here, we drew $13$ random between-clusters connections to keep the clusters joint to each other. In the final and most high-dimensional scenario, the size of the first layer was expanded further to $p_X=130$ nodes, with a reduced connection probability of $\frac{1}{12}$, and $17$ between-cluster random connections. In contrast, the second and smallest layer remained fixed across all scenarios as a two-clusters network with $p_Z=20$ nodes. Here, the probability of within-cluster connections was set to $0.35$, and we joined the two clusters with $4$ random connections. In each scenario, once the structure of the network of both layers was simulated, we used the \textit{huge} data-generating toolkit to generate the respective precision matrix of both networks, which we fixed, respectively, as the diagonal blocks $\boldsymbol{\Theta}_{XX}$ and $\boldsymbol{\Theta}_{ZZ}$ of the target precision matrix. As a consequence, the three scenarios have an increasing dimensionality, namely with $p_X + p_Z = 60$ in the first scenario, $p_X + p_Z = 100$ in the second, and $p_X + p_Z = 150$ in the third.

For all scenarios, the next step of the multi-omics GGM generation was to generate $\boldsymbol{\Theta}_{XZ}$, the off-diagonal block of the precision matrix representing the \textit{between}-layer connections. This step was divided into two main tasks. First, in each scenario, we separately sampled data form the two multivariate normal distributions $X \sim MVN_{p_{X}}(0_{p_{X}},\boldsymbol{\Theta}_{XX}^{-1})$ and $Z \sim MVN_{p_{Z}}(0_{p_{Z}},\boldsymbol{\Theta}_{ZZ}^{-1})$, drawing $100$ observations from each distribution to generate datasets $\boldsymbol{X}$ and $\boldsymbol{Z}$, respectively. Second, we performed a multivariate regression $\boldsymbol{Z} = \boldsymbol{B}\boldsymbol{X} + \boldsymbol{E}$, where $\boldsymbol{B}$ is a $p_Z \times p_X$ coefficients matrix and $\boldsymbol{E}$ is a random error matrix with columns distributed as $E_i  \sim MVN_{p_{Z}}(0_{p_{Z}} ,\boldsymbol{\Theta}_{ZZ}^{-1})$. To perform this multivariate regression we used the R package \textit{MRCE}, which guarantees to compute a sparse $\boldsymbol{B}$ matrix of coefficients. The package uses two sparsity inducing parameters, $\lambda_1$ and $\lambda_2$. We explored a grid of possible combinations of the two, going from a stronger to a weaker penalization. We proceeded on this path, gradually activating a larger proportion of coefficients in the matrix $\boldsymbol{B}$. Once we reached a percentage of activated coefficient as close as possible to $40\%$, we selected the resulting $\boldsymbol{B}$. We then set $\boldsymbol{\Theta}_{ZX} = (\boldsymbol{\Theta}_{XZ})^t = \boldsymbol{B}$. This procedure resulted in off-diagonal blocks with approximately $42\%$ non-zero entries in scenario $1$, $40\%$ in scenario $2$, and $39\%$ in scenario $3$.

Once we had assembled $\boldsymbol{\Theta}$, we guaranteed its positive semi-definitiveness by adding a small constant to its diagonal elements, $\epsilon = 0.3$ in the first two scenarios, and $\epsilon = 0.4$ in the third. In each scenario, we fixed the resulting $\boldsymbol{\Theta}$ as the true precision matrix of the simulated GGM. To conclude the generation of the multi-omics GGM of the three scenarios, we inverted each matrix $\boldsymbol{\Theta}$ to obtain the corresponding $\boldsymbol{\Sigma}$, the variance-covariance matrix of the simulated GGM. 

Once we had generated the GGMs of each scenario, we proceeded to the second major phase of the simulations: generating the simulated multi-omics datasets based on the true $\boldsymbol{\Sigma}$. For each scenario, we sampled data from a multivariate normal distribution with mean $\mu = 0$ and covariance $\boldsymbol{\Sigma}$. In particular, we generated $100$ datasets, or replicates, for each scenario, each one having a total of $n=50$ observations. As a result, we had three increasingly ``high-dimensional'' scenarios, with $n$ fixed to $50$ and $p_X + p_Z$ being, respectively, $60$ in the first scenario, $100$ in the second, and $150$ in the third, each one with an associated complex ground-truth network and its corresponding $\boldsymbol{\Theta}$ matrix.

We applied both \textit{coglasso} and \textit{glasso} to the simulated datasets from the three scenarios to compare their performance. Both \textit{coglasso} and \textit{glasso} need hyperparameters to be specified, so, for \textit{coglasso}, we explored a grid of $10$ $\lambda_w$ and $10$ $\lambda_b$ values, and $c$ values $\{0, 0.1, 0.5, 1, 10\}$, while for \textit{glasso} we explored $20$ $\lambda$ values. The comparison focused on two aspects: the performance of the two methods in reconstructing the structure of the true network; and the fidelity of the two methods in estimating the values of the precision matrix $\widehat{\boldsymbol{\Theta}}$. To measure the performance of the network structure reconstruction, we used the $F_1$ coefficient and the Matthews correlation coefficient (\emph{MCC}), while to measure the quality of the estimated precision matrix we used the Kullback-Leibler divergence (\emph{KLD}). Below, we describe how of each metrics is defined.

For the $F_1$ and \emph{MCC} of a reconstructed network, let us take off-diagonal elements of the adjacency matrix of the true network $\boldsymbol{A}$ and the one of the estimated network $\widehat{\boldsymbol{A}}$, and use them to compute the true positives (\emph{TP}s), false positives (\emph{FP}s), true negatives (\emph{TN}s) and false negatives (\emph{FN}s). We define the $F_1$ score as 
\begin{equation} \label{F1}
    F_1  = 2 \frac{precision \cdot recall}{precision + recall},
\end{equation}
with $precision = TP/(TP + FP)$ and $recall = TP/(TP + FN)$. The $F_1$ score summarizes the balance between precision and recall in identifying the true edges of the graph.
The \emph{MCC}, instead, is generally regarded as a good overall measure that encompasses all the four categories, and it is defined as:
\begin{equation} \label{MCC}
    MCC  = \frac{TP \cdot TN - FP \cdot FN}{\sqrt{(TP + FP)(TP + FN)(TN + FP)(TN + FN)}}.
\end{equation}

The \emph{KLD} measures the distance between the probability distribution of the estimated GGM and the true GGM, and is used to compute the correctness of the values of the estimated precision matrix. For $\boldsymbol{\Theta}$ and $\widehat{\boldsymbol{\Theta}}$ being, respectively, the true and the estimated precision matrix, we formulate it as given by \citet{pardo_statistical_2018}: 
\begin{equation} \label{KLD}
    KLD \left (\boldsymbol{\Theta}, \widehat{\boldsymbol{\Theta}} \right ) = \frac{1}{2} \left ( tr(\boldsymbol{\Theta}\widehat{\boldsymbol{\Theta}}^{-1}) + tr(\widehat{\boldsymbol{\Theta}}\boldsymbol{\Theta}^{-1})  - (p_X+p_Z)\right ).
\end{equation}
 
\begin{figure}[!p]
    \centering
    \includegraphics[width=1\linewidth]{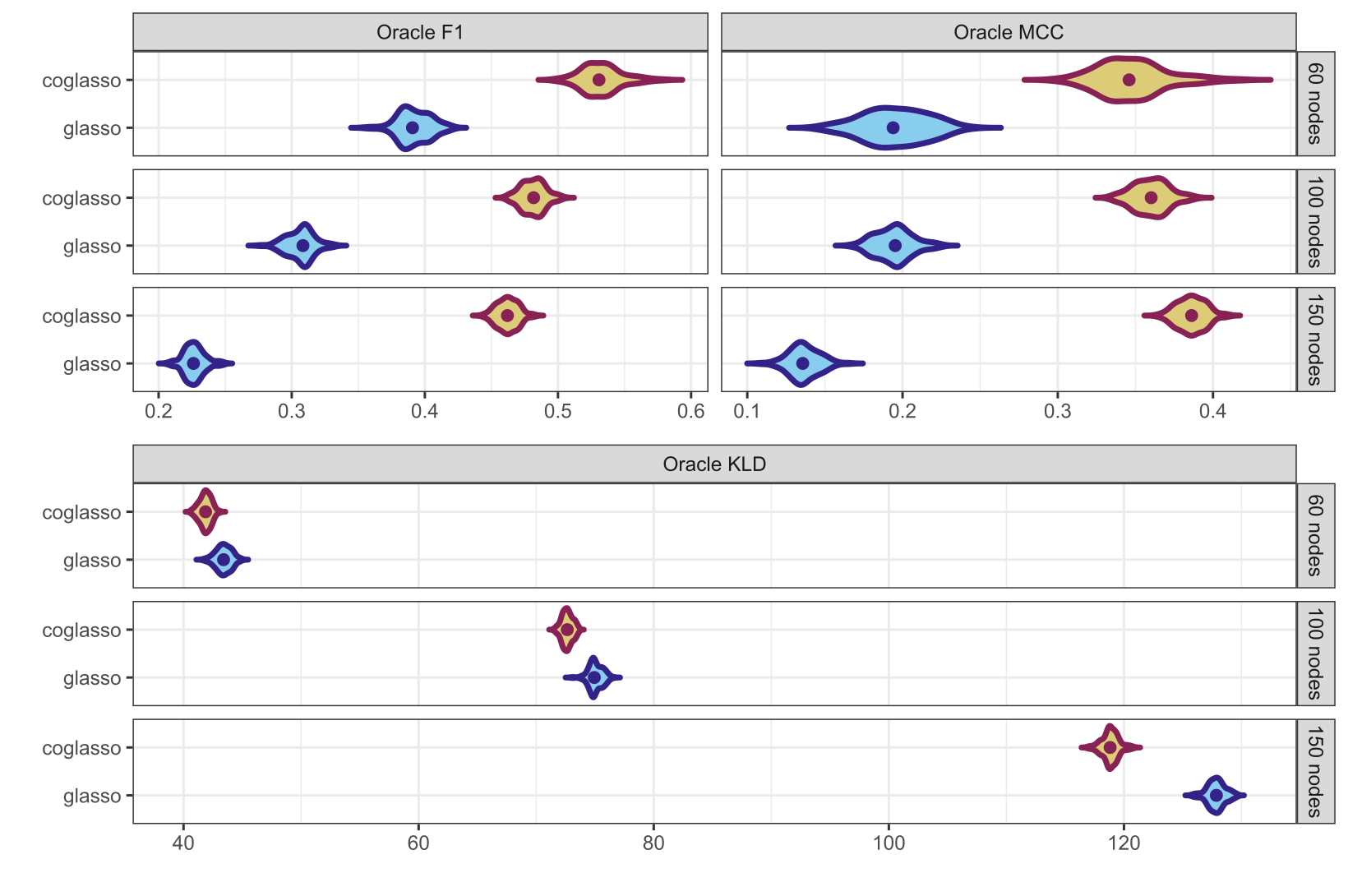}
    \caption{Results of simulated networks reconstruction for \textit{coglasso} (in yellow-red) and \textit{glasso} (in cyan-blue). The performance was measured in terms of network structure recovery (with $F_1$ and \emph{MCC}) and of estimation of the precision matrix (with Kullback-Leibler divergence), in three increasingly complex and high-dimensional scenarios (networks with $60$, $100$, and $150$ nodes). The panel above shows the results when measuring the $F_1$ and the \emph{MCC}, while the one below shows the results when measuring the \emph{KLD}. In each panel the scenarios are distributed along the rows. For each replicate and in all scenarios, the ``oracle'' measure of each method was taken. This means that the figure reports only the measure of the best achieving network of each method from the grid of explored hyperparameters. \textit{Coglasso} shows an advantage over \textit{glasso} according to all measures. The advantage becomes larger as the complexity of the problem increases. Importantly, the $c$ value of the best achieving network of \textit{coglasso} is larger than zero across all replicates of the two most complex scenarios, and only once for the least complex (see Supplementary Figure S1). This implies a role of collaboration in achieving the best possible network. The violins of the figure are composed of $100$ data points, one for each replicate of the simulations, and the networks were reconstructed from datasets of $n = 50$ observations.}
    \label{fig:oracle}
\end{figure}

 Figure \ref{fig:oracle} reports the measured performance of \textit{coglasso} and \textit{glasso}, across all three simulation scenarios, each evaluated over $100$ replicates. For each replicate, the metric was computed for the oracle network, that is the best-performing network selected from the grid of hyperparameter combinations. The three scenarios are distributed along rows of the figure, while the three measures along the columns. In all scenarios \textit{coglasso} consistently outperforms \textit{glasso}, both in terms of network structure recovery and in terms of estimation of the precision matrix, with the oracle $F_1$ coefficients and \emph{MCC}s being consistently higher than \textit{glasso}'s in all scenarios and the oracle \emph{KLD} values being consistently lower. In particular, according to all measures, the advantage of \textit{coglasso} over \textit{glasso} increases as the complexity of the problem increases. Remarkably, across all scenarios and replicates, apart from a single replicate in the least complex scenario, and for all measures, the best achieving \textit{coglasso} network was always one for a $c>0$ (Supplementary Figure S1). This means that collaboration actively contributed to the achievement of the best possible network across all scenarios and according to both measures of network structure recovery and of quality of the estimated precision matrix.

\begin{figure}[!p]
    \centering
    \includegraphics[width=1\linewidth]{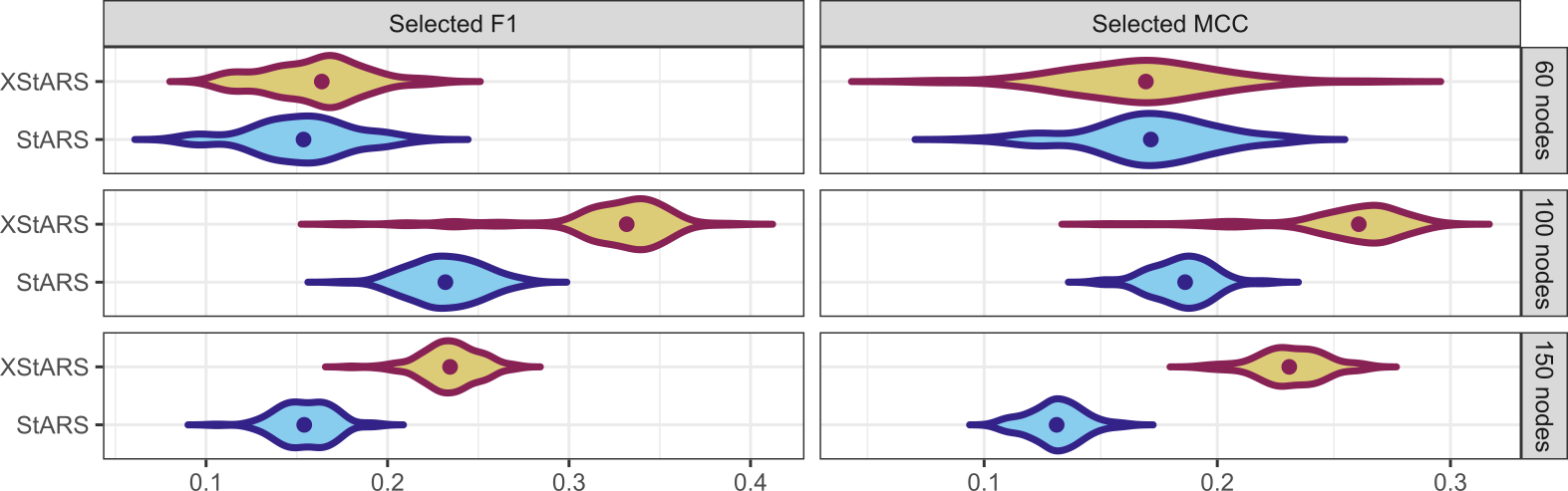}
    \caption{Performance of model selection with \textit{XStARS} for \textit{coglasso} (in yellow-red) and \textit{StARS} for \textit{glasso} (in cyan-blue) over three increasingly complex and high-dimensional scenarios (networks with $60$, $100$, and $150$ nodes). The model selection performance was measured comparing the selected network structure in each replicate with the network structure of the simulated ground truth networks in terms of $F_1$ and \emph{MCC}. The two measures are distributed along the columns of the grid, and the scenarios along the rows. According to the \textit{XStARS} selected \textit{coglasso} network has an especially large advantage over the \textit{StARS} selected \textit{glasso} network in the two most complex scenarios (row two and three). The violins of the figure are composed of $100$ data points, one for each replicate of the simulations, and the networks were reconstructed from datasets of $n = 50$ observations.}
    \label{fig:selected}
\end{figure}

We employed the same set of simulated scenarios used above to test the performance of \textit{XStARS} as a model selection for \textit{coglasso}. In particular, we compared the performance of \textit{XStARS} to that of the closest available method for \textit{glasso}, the original \textit{StARS} algorithm by \citet{liu_stability_2010}. As both methodologies are based on the stability of the edge composition only, without considering the estimated values of the precision matrix, in Figure \ref{fig:selected} we focus our comparison only on the $F_1$ and \emph{MCC} measures of the selected network for each replicate. The figure shows the advantage of \textit{XStARS} in selecting the network with the best structure over \textit{StARS}. This advantage becomes more pronounced as the complexity of the problem increases, particularly in the two most challenging scenarios involving networks with $100$ and $150$ nodes. Supplementary Figure S2 also shows the performance in terms of \emph{KLD}. 

\subsection{Application to a multi-omics study of sleep deprivation in mouse}\label{sec5}

We illustrate the proposed method using a multi-omics dataset from \citet{diessler_systems_2018} to study the molecular biology behind sleep deprivation (SD). This study compares transcriptomic and metabolomic measurements in sleep-deprived mice with non-sleep-deprived mice. The dataset we selected consists of a sample of $30$ diverse mouse lines, $p_x=14896$ genes measured in the cortex and $p_z=124$ blood circulating metabolites, measured in the SD condition. The original study also performs differential expression analysis, providing a list of differentially expressed genes and differentially expressed metabolites. Moreover, the authors mention a list of $78$ genes known to be associated with SD from \citet{mongrain_separating_2010}. We decided to use this information to reduce the dimensionality of the two omics layers. From the transcriptomic layer, we extracted the union of the $78$ known genes and the top $100$ differentially expressed genes, resulting in a list of $162$ genes. Similarly, from the metabolomic layer we extracted the $76$ differentially expressed metabolites. We used \textit{coglasso} to estimate the network from the two aggregated omics layers. 
We run our method over $20$ possible $\lambda_b$ and $\lambda_w$ values, and over $6$ $c$ values spanning from $0$ to $100$, leading to a total of $2400$ combinations of hyperparameters. 
Among the possible networks of the grid, \textit{XStARS} selected the network with hyperparameters $\widehat{\lambda}_w =0.64$, $\widehat{\lambda}_b =0.36$ and $\widehat{c}=0.10$. In Supplementary Figure S3 we show the resulting \textit{coglasso} network.

\begin{figure}[!p]
    \centering
    \includegraphics[width=1\linewidth]{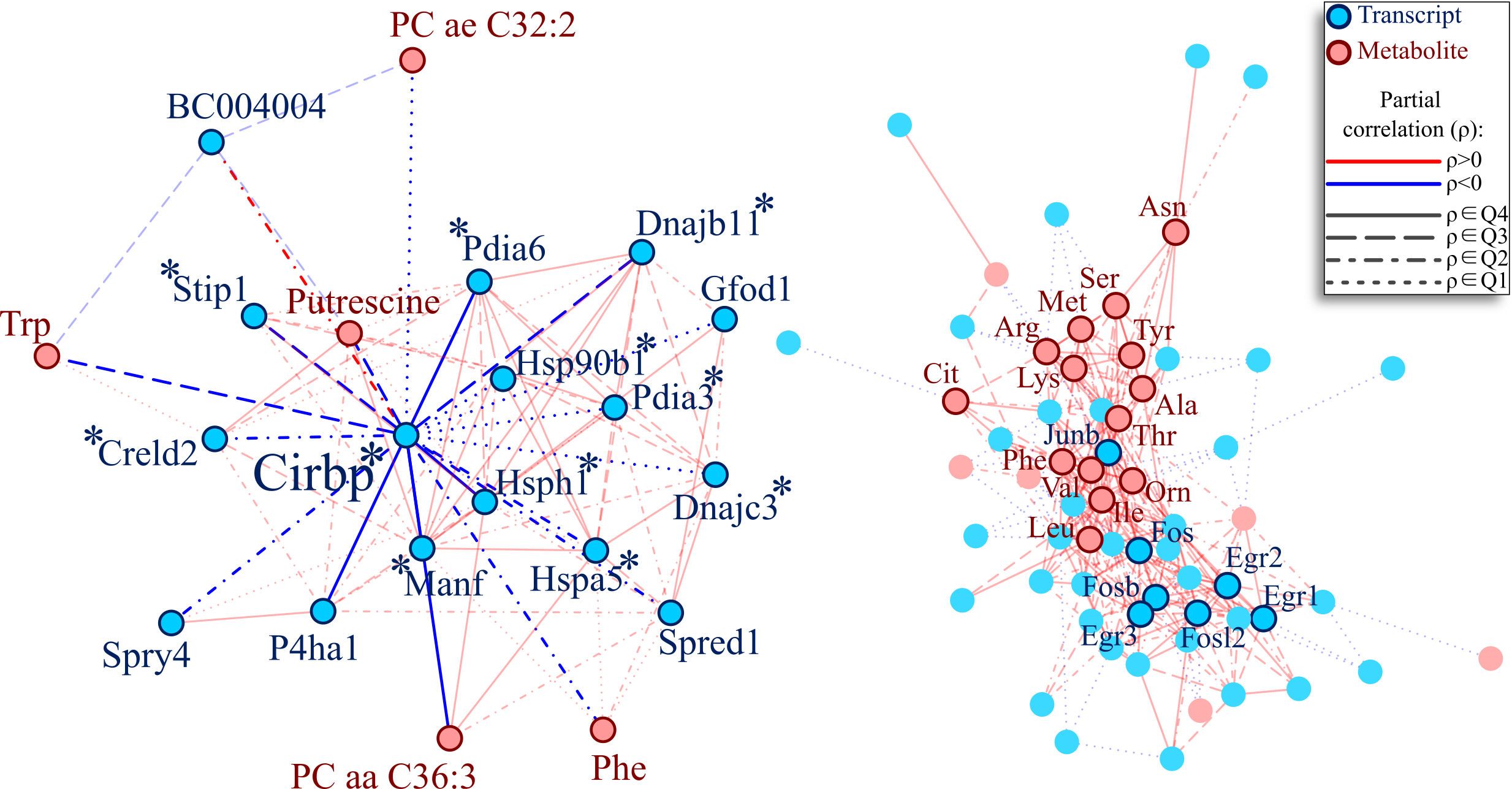}
    \caption{Subnetwork of \textit{Cirbp} and its neighbouring nodes (left) and second largest community (right) from the \textit{coglasso} network shown in Supplementary Figure 3. Blue nodes represent transcripts, while pink nodes represent metabolites. Blue edges represent negative partial correlations, while red edges stand for positive partial correlations. There are four line intensities, representing the strength of the edges. Dotted lines represent the first quartile of edge strengths of the network, while full lines represent the last quartile. On the left, node labels with asterisks belong to genes known to be involved with unfolded protein response or protein folding in general. \textit{Cirbp} shows a negative relation to most of the genes of its neighbourhood.}
    \label{fig:cirbp_comm2}
\end{figure}

To explore the biological meaningfulness of a network generated with \textit{coglasso}, we followed two strategies: the first one knowledge-based and the second one data-driven. In both instances, we were able to show how \textit{coglasso} could retrieve connections that have previously been experimentally validated.\\
For the knowledge-based driven approach, we assessed the available literature on the molecular biology behind sleep. \citet{hoekstra_cold-inducible_2019} investigated the role of Cold-Induced RNA Binding Protein (\textit{Cirbp}) as a molecular regulator of SD response. Figure \ref{fig:cirbp_comm2}, on the left, shows the subnetwork of \textit{Cirbp} and its neighbouring nodes. Among the targets \citet{hoekstra_cold-inducible_2019} investigated, two genes encoding for heat-shock proteins: \textit{Hspa5} and \textit{Hsp90b1}, had an acutely increased induction upon SD in \textit{cirbp} knock-out mutant mice. This indicates that these two genes act in response to SD under the regulation of \textit{Cirbp}. In the selected \textit{coglasso} network, \textit{Cirbp} was connected to both these genes, meaning that \textit{coglasso} was able to reconstruct connections that have previously been validated. These two genes are involved in unfolded protein response (UPR), a process known to be activated during SD \citep{ji_liver-specific_2011,sun_endoplasmic_2019,mackiewicz_macromolecule_2007}. Interestingly, eight additional genes in the neighbourhood of \textit{Cirbp} are associated with UPR or, in general, with protein folding \citep{solda_consequences_2006,vekich_protein_2012,kern_creld2_2021,lin_hsp90_2021,mizobuchi_armet_2007,genereux_unfolded_2015,tao_chapter_2011,yamagishi_characterization_2011}. Hence, \textit{coglasso} estimated biologically coherent connections between \textit{Cirbp} and other UPR-involved genes. This suggests that \textit{Cirbp} could act as a regulator of the UPR upon SD, a hypothesis that could be experimentally tested. Additionally, the connection between \textit{Cirbp} and tryptophan (Trp) suggests a potential regulatory role in response to SD, since Trp is a precursor to serotonin, one of the key hormones in sleep regulation \citep{minet-ringuet_tryptophan-rich_2004}.

We then inspected the generated network by a data-driven approach, performing community discovery with the algorithm described by \citet{clauset_finding_2004}. We focused on the second-largest community that the community discovery algorithm identified (shown in Figure \ref{fig:cirbp_comm2}, on the right, the community is highlighted in the global network in Figure S4). Among other metabolites, the second community contained the majority of the amino acids recorded in the dataset: alanine, arginine, asparagine, citrulline, isoleucine, leucine, lysine, methionine, ornithine, phenylalanine, serine, threonine, tyrosine and valine. Among the transcripts, it included all the genes belonging to the \textit{Fos Proto-Oncogene} (\textit{Fos})/\textit{Jun Proto-Oncogene} (\textit{Jun}) and the \textit{Early Growth Response} (\textit{Egr}) transcription factor families that were present in our network: \textit{Fos}, \textit{Fosb}, \textit{Fosl2}, \textit{Junb}, \textit{Egr1}, \textit{Egr2}, and \textit{Egr3}. Their nodes are those highlighted in the figure. Several previous studies reported that various members of the two transcription factor families participate in the amino acid starvation response (AAR, \citep{kilberg_transcription_2012}. For example, transcriptomic studies in mouse \citep{deval_amino_2009} and human cells \citep{shan_expression_2010} showed induction of members of both transcription factor families. In particular, in \citet{deval_amino_2009},
\textit{Fos} was the most upregulated member of the \textit{Fos/Jun} transcription factor family upon leucine starvation, a connection that \textit{coglasso} recovered.
Previously, \citet{pohjanpelto_deprivation_1990} separately studied the induction of targeted members of the \textit{Fos/Jun} family to methionine deprivation in hamster cells, with \textit{Junb} showing again the strongest response. This connection is present in the community too. Moreover, despite histidine is not part of the community identified by the clustering algorithm, \textit{coglasso} connected it to \textit{Junb}, and in \citet{shan_expression_2010}, \textit{Junb} exhibited the greatest induction upon histidine deprivation. Other targeted studies verified the role during AAR of \textit{Egr1} \citep{shan_mitogen-activated_2014,shan_induction_2019} and \textit{Fos} \citep{shan_mapk_2015} in human cells. Since the two transcription factor families belong to the group of immediate-early response genes and take part in AAR, their expression is expected to fluctuate with high sensitivity accordingly with variations in amino acid concentrations \citep{healy_immediate_2013}. Altogether, these data show that \textit{coglasso} was able to reconstruct several previously validated connections between transcripts and amino acids, as well as estimating additional biologically coherent links between variables of these two layers.

\section{Discussion}
\label{sec6}
In this study, we addressed the issue of estimating a network from multiple high-dimensional datasets of variables that are different in nature and are recorded on the same individuals (\textit{e.g.} multi-omics datasets). In particular, we focused on GGM estimation in a multi-layer setting. To tackle this question, we introduced \textit{coglasso}, a new algorithm that is able to leverage the contribution of diverse omics layers in the process of GGM estimation. \textit{Coglasso} is based on major changes in \textit{glasso} \citep{friedman_sparse_2008}, a well-known GGM estimation algorithm. The cornerstone of this alteration was the introduction of a collaborative term in the objective function, similar to what \citet{gross_collaborative_2015} described with \textit{collaborative regression}. Along with it, we also introduced a mixed $L_1$-penalty system that separately promotes sparsity \textit{within} and \textit{between} omics layers with two hyperparameters $\lambda_w$ and $\lambda_b$. With the introduction of a collaborative term, coming with its own hyperparameter $c$, and of the mixed $L_1$-penalty system, it became necessary to develop an appropriate model selection strategy for \textit{coglasso}. One of the best approaches to address this for \textit{glasso}, based on the network stability criterion, is \textit{StARS} \citep{liu_stability_2010}. Nevertheless, this strategy was designed to handle a single hyperparameter. Therefore, we developed \textit{eXtended }\textit{StARS} (\textit{XStARS}) to cope with the additional hyperparameters. 

After designing and implementing \textit{coglasso} and \textit{XStARS}, we then compared network reconstruction performances of \textit{coglasso} and \textit{glasso} by simulation studies. The comparisons showed an advantage of \textit{coglasso} over \textit{glasso}. In particular, this advantage increased along with the complexity of the simulated multi-omics scenario.
Our simulation studies also showed that the collaborative term actively influences the network reconstruction performance, contributing to the best network that \textit{coglasso} can achieve. Additionally, in our simulations \textit{XStARS}-selected \textit{coglasso} networks outperformed \textit{StARS}-selected \textit{glasso} networks, showing an advantage at the level of model selection too. We note here that, although this simulation study aimed to reflect a range of structural complexities, real-world multi-omics data often exhibit intricate dependencies that are difficult to reproduce in current simulation frameworks. Therefore, we look forward to testing \textit{coglasso} on more biologically realistic simulated datasets as such benchmarks become available. It is noteworthy to mention that despite the methodological similarity, we were unable to include \textit{DRAGON} in our simulation study due to the lack of publicly available documentation. Additionally, popular multi-omics data integration tools like \textit{MOFA} and \textit{DIABLO} could not be included either. This is because their main objectives are, respectively, low-dimensional representation of multi-omics data and multi-omics biomarker identification, while \textit{coglasso}'s main objective is multi-omics network reconstruction.

Following the simulations study, we illustrated \textit{coglasso}’s ability to reconstruct networks from real-world multi-omics data. We applied the method to a multi-omics dataset studying transcriptomics and metabolomics of sleep deprivation in mice \citep{hoekstra_cold-inducible_2019}. \textit{Coglasso}, together with our novel model selection procedure, was able to retrieve connections that had previously been validated, both surrounding the gene \textit{Cirbp} and in the community of the network most enriched with amino acids. Moreover, the first example allowed the formulation of additional hypotheses on the regulatory behaviour of the gene \textit{Cirbp} in sleep deprivation, and we encourage the field of sleep studies to assess these on biological relevance.

Our results show that \textit{coglasso} represents a novel relevant alternative in the quest for multi-omics network reconstruction, as it uniquely leverages the multi-layer nature of multi-omics data. 
Nevertheless, like many other GGM estimation algorithms, \textit{coglasso} relies on the normality assumption, which is rarely true for multi-omics data. In recent years, researchers have attempted to solve this major issue by resorting to copula-based techniques \citep{behrouzi_detecting_2019,cougoul_magma_2019,vinciotti_bayesian_2022}. Therefore, extending our algorithm with a copula-based approach would allow the transition of multi-omics data to the normal realm, and, by doing so, could generalize the applicability of \textit{coglasso}.

While the implementation of \textit{coglasso} in the current R package supports the integration of two omics layers, multi-omics studies could generate more than two types of data. Extending \textit{coglasso} to accommodate multiple omics layers is a promising direction for future development.
Additionally, where \textit{coglasso} integrates multiple layers of variables, tools like \textit{fused graphical lasso} \citep{danaher_joint_2014} allow to integrate multiple experimental conditions in a single network. Hence, extending \textit{coglasso} to include a \textit{fused graphical lasso}-type penalty would be worthwhile pursuing.

Finally, even though we designed \textit{coglasso} as an effective strategy to integrate multi-omics data, it remains a general method. This implies that \textit{coglasso} has a wide margin of applicability in any scientific field that could be interested in reconstructing networks leveraging datasets of multiple layers of variables. For example, in the field of neuroscience, it is common to associate different functions to different anatomical areas of the brain \citep{kanwisher_functional_2010}. Therefore, \textit{coglasso} could contribute to the reconstruction of neural connectivity networks from functional Magnetic Resonance Imaging data, defining different brain areas as separate layers of variables.

\paragraph*{Competing interests: no competing interest is declared.}

\section*{Author contributions statement}
Alessio Albanese (Conceptualization [equal], Data curation [lead], Formal analysis [lead], Investigation [lead], Methodology [lead], Software [lead], Validation [lead], Visualization [lead], Writing—original draft [lead], Writing—review \& editing [equal]), 
Wouter Kohlen (Conceptualization [supporting], Funding acquisition [equal], Investigation [supporting], Supervision [supporting], Writing—review \& editing [supporting]), 
and Pariya Behrouzi (Conceptualization [equal], Formal analysis [supporting], Funding acquisition [equal],  Investigation [supporting], Methodology [supporting], Software [supporting], Supervision [lead], Writing—original draft [supporting], Writing—review \& editing [equal]).

\section*{Acknowledgments}
We thank Prof. Fred van Eeuwijk and Dr. Gwena\"{e}l G.R. Leday (Wageningen University and Research) for fruitful discussions during the development of \textit{coglasso}. We thank Dr. Michael Schon (Wageningen University and Research) for the valuable intellectual discussions on the biological application. 

\section*{Funding}
This work was funded by the Plant Science Group, the departments of Mathematical and Statistical Methods (Biometris) and of Cell and Developmental Biology at Wageningen University and Research [project number 3183500140]; and by the NWO VIDI [grant number VI.Vidi.193.119] (W.K.).

\bibliographystyle{plainnat}
\bibliography{coglasso}

%\begin{figure}[!p]
%\centering\includegraphics[width=1.1\textwidth]{selected_F1_MCC.png}
%\caption{Performance of model selection with \textit{XStARS} for \textit{coglasso} (in yellow-red) and \textit{StARS} for \textit{glasso} (in cyan-blue) over three increasingly complex and high-dimensional scenarios (networks with $60$, $100$, and $150$ nodes). The model selection performance was measured comparing the selected network structure in each replicate with the network structure of the simulated ground truth networks in terms of $F_1$ and \emph{MCC}. The two measures are distributed along the columns of the grid, and the scenarios along the rows. According to the \textit{XStARS} selected \textit{coglasso} network has an especially large advantage over the \textit{StARS} selected \textit{glasso} network in the two most complex scenarios (row two and three). The violins of the figure are composed of $100$ data points, one for each replicate of the simulations, and the networks were reconstructed from datasets of $n = 50$ observations.}
%\label{fig:selected}
%\end{figure}

%\begin{figure}[!p]
%\centering\includegraphics[width=\textwidth]{cirbp_comm2.png}
%\caption{Subnetwork of \textit{Cirbp} and its neighbouring nodes (left) and second largest community (right) from the \textit{coglasso} network shown in Figure \ref{fig:network}. Blue nodes represent transcripts, while pink nodes represent metabolites. Blue edges represent negative partial correlations, while red edges stand for positive partial correlations. There are four line intensities, representing the strength of the edges. Dotted lines represent the first quartile of edge strengths of the network, while full lines represent the last quartile. On the left, node labels with asterisks belong to genes known to be involved with unfolded protein response or protein folding in general. \textit{Cirbp} shows a negative relation to most of the genes of its neighbourhood.}
%\label{fig:cirbp_comm2}
%\end{figure}

\end{document}

% Title of paper
\title{Collaborative graphical lasso supplementary material}

\author[1,2]{Alessio Albanese}
\author[2]{Wouter Kohlen}
\author[1]{Pariya Behrouzi*}
\affil[1]{Mathematical and Statistical Methods (Biometris),
Wageningen University and Research}
\affil[2]{Cellular and Developmental Biology,
Wageningen University and Research}
\date{}
\renewcommand\Affilfont{\itshape\small}
\renewcommand*{\thefootnote}{\fnsymbol{footnote}}

\maketitle

% Add a footnote for the corresponding author if one has been
% identified in the author list
\footnotetext{\textsuperscript{*}To whom correspondence should be addressed. Email: pariya.behrouzi@wur.nl}

\section{Algorithms}
\subsection{Update rule of the coordinate descent algorithms of \textit{glasso}}
The coordinate descent procedure described in \citet{friedman_pathwise_2007} estimates $\boldsymbol{\widehat{\beta}_i}$ from Equation 1 of the main text, for any $i$, by iteratively updating the $j$-th coordinate of the vector, $(\boldsymbol{\widehat{\beta}_i})_j$, which is the main proxy for the conditional dependence between node $i$ and node $j$ of the network.
\begin{algorithm}
\caption{\textit{glasso} coordinate descent rule}\label{alg:glasso_cdr}
\begin{algorithmic}
\For{$j=1,2,\ldots,p,1,2,\ldots,p,\ldots$, until convergence criterion is met}
\State $r\gets \boldsymbol{S}_{ji}-\sum_{k\neq j} \left ( \boldsymbol{W}_{\setminus i\setminus i} \right )_{jk}\left ( \boldsymbol{\widehat{\beta}_{i}} \right )_k$
\If{$r<-\lambda$}
    \State $\left ( \boldsymbol{\widehat{\beta}_i} \right )_j\gets \frac{r+\lambda}{\boldsymbol{W}_{jj}}$
\ElsIf{$r>\lambda$}
    \State $\left ( \boldsymbol{\widehat{\beta}_i} \right )_j\gets \frac{r-\lambda}{\boldsymbol{W}_{jj}}$
\Else
    \State $\left ( \boldsymbol{\widehat{\beta}_i} \right )_j\gets 0$
\EndIf
\EndFor
\end{algorithmic}
\end{algorithm}

The Algorithm begins with computing an initial value $r$. The value assigned to $r$ is computed using two terms. The first one is the entry $\boldsymbol{S}_{ji}$ of the empirical variance-covariance matrix, the common variability of node $i$ and node $j$. This common variability is adjusted for the role that all the other nodes $k$ of the network exert on node $i$ and on node $j$ by subtracting a dot product to $\boldsymbol{S}_{ji}$. This reflects the principles of conditional dependence described above. After $r$ is computed, its value can either fall outside or inside of the penalty interval $(-\lambda, \lambda)$. In the first case, the algorithm assigns to $ (\boldsymbol{\widehat{\beta}_i})_j$ the value of $r$ shrunk towards the interval. In the other, $(\boldsymbol{\widehat{\beta}_i})_j$ is fixed to $0$. The updates are run multiple times for $j=1,2,\ldots,p,1,2,\ldots,p,\ldots$ until convergence of the vector $\boldsymbol{\widehat{\beta}_i}$.

\subsection{Update rule of the coordinate descent algorithms of \textit{coglasso}}
The chosen algorithm to solve the objective function of \textit{collaborative graphical lasso} shown in Equation 2 of the main text is also a coordinate descent procedure. Here follows the update rule of the procedure, used by \textit{coglasso} to estimate the $j$-th element of $\boldsymbol{\beta_i}$, which represents a proxy for the connection between the $i$-th and the $j$-th variable. For the details about the derivation, we refer the reader to the Appendix A of these Supplementary Data. To keep the update rule more general, let $j$ belong to a generic indexes set $A$, which could, for instance, represent variables from the omics layer $\boldsymbol{X}$.
\begin{algorithm}
\caption{\textit{coglasso} coordinate descent rule}\label{alg:coglasso_cdr}
\begin{algorithmic}
\For{$j=1,2,\ldots,p,1,2,\ldots,p,\ldots$, until convergence criterion is met}
\State $r \gets \alpha \boldsymbol{S}_{ji} -\sum_{k\in A, k\neq j} \left ( \boldsymbol{W}_{\setminus i\setminus i} \right )_{jk}\left ( \boldsymbol{\widehat{\beta}_{i}} \right )_k - \alpha(1-c)\sum_{k\notin A} \left ( \boldsymbol{W}_{\setminus i\setminus i} \right )_{jk}\left ( \boldsymbol{\widehat{\beta}}_{i} \right )_k$ 
\If{$r<-\alpha \boldsymbol{\Lambda}_{ji}$}
    \State $\left ( \boldsymbol{\widehat{\beta}}_i \right )_j\gets \frac{r+\alpha \boldsymbol{\Lambda}_{ji} }{\boldsymbol{W}_{jj}}$
\ElsIf{$r>\alpha \boldsymbol{\Lambda}_{ji}$}
    \State $\left ( \boldsymbol{\widehat{\beta}}_i \right )_j\gets \frac{r-\alpha \boldsymbol{\Lambda}_{ji} }{\boldsymbol{W}_{jj}}$
\Else
    \State $\left ( \boldsymbol{\widehat{\beta}_i} \right )_j\gets 0$
\EndIf
\EndFor
\end{algorithmic}
\end{algorithm}

Here, $\alpha = \frac{1}{1+c}$. There are remarkable differences with \textit{glasso}’s update rule seen in Algorithm \ref{alg:glasso_cdr} to update $(\boldsymbol{\widehat{\beta}_i})_j$. First, here entry $\boldsymbol{S}_{ji}$ is weighted by $\alpha$. Second, when adjusting for the influence that nodes $k$ exert on the connection between node $i$ and node $j$, \textit{coglasso} differentiates between two groups of nodes: nodes that belong to the same set of $j$ ($k\in A$), and nodes that do not ($k\notin A$). When a node $k$ belongs to the same set as $j$, the adjustment for the role of those nodes always has the same weight and direction, like for \textit{glasso}. When instead it doesn’t, the weight and direction of the adjustment depends on the c parameter. When $c = 0$ there is no difference between \textit{coglasso} and \textit{glasso} in computing $r$, and so in adjusting $\boldsymbol{S}_{ji}$ and the two adjustments for $k\in A$ and $k\notin A$ merge into a single update rule. For $0 < c < 1$, the direction stays the same as the other group of nodes, but the weight of the adjustments slowly decreases. For $c = 1$, the weight of these nodes in adjusting the dependence between node $i$ and node $j$ is $0$. For values of $c$ higher than $1$, the direction of the adjustment for the nodes that do not belong to the same omics layer as $j$ inverts the direction, increasing its weight as the value of $c$ increases. A final difference is that in \textit{coglasso} the penalty interval is now defined as $(-\alpha\boldsymbol{\Lambda}_{ji}, \alpha\boldsymbol{\Lambda}_{ji})$. Then, similarly to \textit{glasso}, if $r$ falls out of it, we assign to $(\boldsymbol{\widehat{\beta}_i})_j$ a value closer than $r$ to the interval, while if $r$ falls inside it we assign $0$ to it.

\subsection{eXtended Stability Approach to Regularization Selection (\textit{XStARS})}\label{sec2}
Here we report our generalization of the \textit{StARS} model selection procedure developed in \citet{liu_stability_2010}, \textit{XStARS}. \textit{XStARS}  adapts \textit{StARS} to select hyperparameters in the three-dimensional hyperparameter space of \textit{coglasso}.

Algorithm \ref{alg:xstars_coglasso} describes the outline of \textit{XStARS}, with $\boldsymbol{c}$, $\boldsymbol{\lambda_b}$, and $\boldsymbol{\lambda_w}$ being the arrays of explored values for the hyperparameters. Here, the function $StARS(\boldsymbol{h})$ represents the same one-dimensional algorithm developed for \textit{glasso}. It returns the hyperparameter value $h$ in an array $\boldsymbol{h}$ yielding the most stable, yet sparse \textit{coglasso} network, after fixing the other two hyperparameters to a given value. In other words \textit{XStASRS} alternately estimates an optimal value for $\lambda_w$, $\lambda_b$, or $c$, keeping the other two fixed to the last optimal value obtained. This alternation continues until  one of two conditions is met. The first one is if the algorithm converges to a stable triplet of hyperparameter values $(\widehat{\lambda}_w, \widehat{\lambda}_b, \widehat{c})$ yielding the optimal stable, yet sparse network. In practice, for this happens when twice in a row the selection of two hyperparameters yields the same values it selected in the previous iteration of the algorithm. Alternatively, the alternation continues until the algorithm has cycled over the three hyperparameters for a predetermined maximum number of iterations, yielding the last triplet it selected. We choose as a default $10$ iterations, an empirically chosen value that offers a good trade-off between stability and computational cost.
\begin{algorithm}
\caption{\textit{XStARS}}\label{alg:xstars_coglasso}
\begin{algorithmic}
\State $MAX\_ITER \gets 10$ (as a default value)
\State $converged \gets 0$
\State $\widehat{\lambda}_w \gets -1$
\State $\widehat{\lambda}_b \gets min(\boldsymbol{\lambda_b})$
\State $\widehat{c} \gets Max(\boldsymbol{c})$
\State $select_\_\lambda_w \gets TRUE$
\State $select_\_\lambda_b \gets FALSE$
\State $iter \gets 0$
\While{$converged < 2$ $AND$ $iter<MAX\_ITER$}
\If{$select_\_\lambda_w$}
\State $\lambda_w^{temp} \gets StARS(\boldsymbol{\lambda_w})$, with $\lambda_b \gets \widehat{\lambda}_b$ and $c \gets \widehat{c}$ 
\If{$\lambda_w^{temp} = \widehat{\lambda}_w$}
\State $converged \gets converged + 1$
\Else
\State $converged \gets 0$
\EndIf
\State $\widehat{\lambda}_w \gets \lambda_w^{temp}$
\State $select_\_\lambda_w \gets FALSE$
\State $select_\_\lambda_b \gets TRUE$
\ElsIf{$select_\_\lambda_b$}
\State $\lambda_b^{temp} \gets StARS(\boldsymbol{\lambda_b})$, with $\lambda_w \gets \widehat{\lambda}_w$ and $c \gets \widehat{c}$ 
\If{$\lambda_b^{temp} = \widehat{\lambda}_b$}
\State $converged \gets converged + 1$
\Else
\State $converged \gets 0$
\EndIf
\State $\widehat{\lambda}_b \gets \lambda_b^{temp}$
\State $select_\_\lambda_b \gets FALSE$
\Else
\State $c^{temp} \gets StARS(\boldsymbol{c})$, with $\lambda_w \gets \widehat{\lambda}_w$ and $\lambda_b \gets \widehat{\lambda}_b$
\If{$c^{temp} = \widehat{c}$}
\State $converged \gets converged + 1$
\Else
\State $converged \gets 0$
\EndIf
\State $\widehat{c} \gets c^{temp}$
\State $select_\_\lambda_w \gets TRUE$
\State $select_\_\lambda_b \gets FALSE$
\EndIf
\State $iter \gets iter + 1$
\EndWhile
\State Return $(\widehat{\lambda}_w, \widehat{\lambda}_b, \widehat{c})$
\end{algorithmic}
\end{algorithm}

\section{Dimensionality of the partitions of \texorpdfstring{$\boldsymbol{W}_{\setminus i\setminus i}$}{W}} 
It is relevant to notice that the dimensions of the partitions of $\boldsymbol{W}_{\setminus i\setminus i}$ described in the manuscript depend on the index $i$. For example, we split $\boldsymbol{W}_{\setminus i\setminus i}$, of dimension $(p-1)*(p-1)$, into two rectangular vertical matrixes: $(\boldsymbol{W}_{\setminus i\setminus i})_X$ and $(\boldsymbol{W}_{\setminus i\setminus i})_Z$. Each one of these two vertical sections can have two different dimensions, depending on the position of the index $i$. The vertical section $(\boldsymbol{W}_{\setminus i\setminus i})_X$ has dimensions $(p-1)*p_x$ when the $i$-th row and column of $\boldsymbol{W}$ the algorithm is updating corresponds to a variable that belongs to the omics layer $\boldsymbol{Z}$. Otherwise, this first vertical section has dimensions $(p-1)*(p_x-1)$ if the $i$-th variable belongs to the to the omics layer $\boldsymbol{X}$. Comparably, $(\boldsymbol{W}_{\setminus i\setminus i})_Z$ can either have dimensions $(p-1)*(p_z-1)$ if the $i$-th variable belongs to omics layer $\boldsymbol{Z}$ and $(p-1)*p_z$ if it does not. Analogous reasoning can be applied for the dimension of the two partitions of the coefficient vector $\boldsymbol{\beta_i}$. $(\boldsymbol{\beta_i})_X$ can have either $p_x$ or $p_x-1$ elements, while $(\boldsymbol{\beta_i})_Z$ can have either $p_z$ or $p_z-1$ elements, depending on where the index $i$ is located. This partitioning ensures that the collaborative penalty term in Equation 2 of the main manuscript remains well-defined across all omics layers, aligning dimensional consistency between the estimated coefficient vectors from omics layers $\boldsymbol{X}$ and $\boldsymbol{Z}$.

\clearpage

\section*{Appendix A: Derivation of \textit{coglasso} coordinate descent update rule}

In this appendix, we report the original \textit{coglasso} objective function, we develop it and we minimize it with respect to the $j$th component of the vector $\boldsymbol{\beta_i}$. Here is the objective function:

\begin{dmath*}
    F_{\boldsymbol{\Lambda}, \boldsymbol{W}_{\setminus i\setminus i}}(\boldsymbol{\beta_i}) =  \frac{1}{2}\left \| (\boldsymbol{W}_{\setminus i\setminus i})^{-1/2}\boldsymbol{S}_i-(\boldsymbol{W}_{\setminus i\setminus i})^{1/2}\boldsymbol{\beta_i}\right \|^2
    \\ + \frac{c}{2}\left \| (\boldsymbol{W}_{\setminus i\setminus i})^{1/2}_X(\boldsymbol{\beta_i})_X- (\boldsymbol{W}_{\setminus i\setminus i})^{1/2}_Z(\boldsymbol{\beta_i})_Z\right \|^2 + \left \| \boldsymbol{\Lambda}_i \odot \boldsymbol{\beta}_i \right \|_1
\end{dmath*}

For an explanation of the elements of the objective function, we refer the reader to Sections 2 and 3 of the manuscript. Now, developing the elements of the objective function:
\begin{align*}
    F_{\boldsymbol{\Lambda}, \boldsymbol{W}_{\setminus i\setminus i}}(\boldsymbol{\beta_i}) & = \frac{1}{2}\boldsymbol{S}_i^T(\boldsymbol{W}_{\setminus i\setminus i}^{-1/2})^T\boldsymbol{W}_{\setminus i\setminus i}^{-1/2}\boldsymbol{S}_i + \frac{1}{2}\boldsymbol{\beta_i}^T(\boldsymbol{W}_{\setminus i\setminus i}^{1/2})^T\boldsymbol{W}_{\setminus i\setminus i}^{1/2}\boldsymbol{\beta_i} - \boldsymbol{\beta_i}^T(\boldsymbol{W}_{\setminus i\setminus i}^{-1/2})^T\boldsymbol{W}_{\setminus i\setminus i}^{1/2}\boldsymbol{S}_i \\
    & + \frac{c}{2}(\boldsymbol{\beta_i})_X^T(\boldsymbol{W}_{\setminus i\setminus i}^{1/2})_X^T(\boldsymbol{W}_{\setminus i\setminus i}^{1/2})_X(\boldsymbol{\beta_i})_X + \frac{c}{2}(\boldsymbol{\beta_i})_Z^T(\boldsymbol{W}_{\setminus i\setminus i}^{1/2})_Z^T(\boldsymbol{W}_{\setminus i\setminus i}^{1/2})_Z(\boldsymbol{\beta_i})_Z \\
    & - \frac{c}{2}(\boldsymbol{\beta_i})_X^T(\boldsymbol{W}_{\setminus i\setminus i}^{1/2})_X^T(\boldsymbol{W}_{\setminus i\setminus i}^{1/2})_Z(\boldsymbol{\beta_i})_Z
    - \frac{c}{2}(\boldsymbol{\beta_i})_Z^T(\boldsymbol{W}_{\setminus i\setminus i}^{1/2})_Z^T(\boldsymbol{W}_{\setminus i\setminus i}^{1/2})_X(\boldsymbol{\beta_i})_X
    + \left \| \boldsymbol{\Lambda}_i \odot \boldsymbol{\beta}_i \right \|_1
\end{align*}

Considering that the matrix $\boldsymbol{W}_{\setminus i\setminus i}$ can be divided in the following block matrix:
\begin{equation*}
    \boldsymbol{W}_{\setminus i\setminus i} = \begin{pmatrix}
        (\boldsymbol{W}_{\setminus i\setminus i})_{XX} & (\boldsymbol{W}_{\setminus i\setminus i})_{XZ}\\
(\boldsymbol{W}_{\setminus i\setminus i})^T_{XZ} & (\boldsymbol{W}_{\setminus i\setminus i})_{ZZ}
    \end{pmatrix}
\end{equation*}

and taking into account the equivalences shown in Appendix B, it's easy to see what follows:

\begin{itemize}
    \item $(\boldsymbol{W}_{\setminus i\setminus i}^{-1/2})^T\boldsymbol{W}_{\setminus i\setminus i}^{-1/2} = \Theta_{\setminus i\setminus i}$
    \item $(\boldsymbol{W}_{\setminus i\setminus i}^{1/2})_X^T(\boldsymbol{W}_{\setminus i\setminus i}^{1/2})_X = (\boldsymbol{W}_{\setminus i\setminus i})_{XX}$ and $(\boldsymbol{W}_{\setminus i\setminus i}^{1/2})_Z^T(\boldsymbol{W}_{\setminus i\setminus i}^{1/2})_Z = (\boldsymbol{W}_{\setminus i\setminus i})_{ZZ}$
    \item $(\boldsymbol{W}_{\setminus i\setminus i}^{1/2})_X^T(\boldsymbol{W}_{\setminus i\setminus i}^{1/2})_Z = (\boldsymbol{W}_{\setminus i\setminus i})_{XZ}$ and $(\boldsymbol{W}_{\setminus i\setminus i}^{1/2})_Z^T(\boldsymbol{W}_{\setminus i\setminus i}^{1/2})_X = (\boldsymbol{W}_{\setminus i\setminus i})^T_{XZ}$
\end{itemize}

Considering the previous rewritings, the objective function can be simplified to the following form:
\begin{align*}
    F_{\boldsymbol{\Lambda}, \boldsymbol{W}_{\setminus i\setminus i}}(\boldsymbol{\beta_i}) & = \frac{1}{2}\boldsymbol{S}_i^T\boldsymbol{\Theta}_{\setminus i\setminus i}\boldsymbol{S}_i + \frac{1}{2}\boldsymbol{\beta_i}^T\boldsymbol{W}_{\setminus i\setminus i}\boldsymbol{\beta_i} - \boldsymbol{\beta_i}^T\boldsymbol{S}_i + \frac{c}{2}(\boldsymbol{\beta_i})_X^T(\boldsymbol{W}_{\setminus i\setminus i})_{XX}(\boldsymbol{\beta_i})_X\\
    & + \frac{c}{2}(\boldsymbol{\beta_i})_Z^T(\boldsymbol{W}_{\setminus i\setminus i})_{ZZ}(\boldsymbol{\beta_i})_Z - \frac{c}{2}(\boldsymbol{\beta_i})_X^T(\boldsymbol{W}_{\setminus i\setminus i})_{XZ}(\boldsymbol{\beta_i})_Z\\
    & - \frac{c}{2}(\boldsymbol{\beta_i})_Z^T(\boldsymbol{W}_{\setminus i\setminus i})^T_{XZ}(\boldsymbol{\beta_i})_X + \left \| \boldsymbol{\Lambda}_i \odot \boldsymbol{\beta}_i \right \|_1
\end{align*}

We now define the matrix of off-diagonal blocks obtained from $\boldsymbol{W}_{\setminus i\setminus i}$ as:
\begin{equation*}
    \boldsymbol{W}_{\setminus i\setminus i}^{ODB} = \begin{pmatrix}
        \boldsymbol{0}_{XX} & (\boldsymbol{W}_{\setminus i\setminus i})_{XZ}\\
(\boldsymbol{W}_{\setminus i\setminus i})^T_{XZ} & \boldsymbol{0}_{ZZ}
    \end{pmatrix}
\end{equation*}

It's now straightforward to see that
\begin{equation*}
    (\boldsymbol{\beta_i})_X^T(\boldsymbol{W}_{\setminus i\setminus i})_{XZ}(\boldsymbol{\beta_i})_Z + (\boldsymbol{\beta_i})_Z^T(\boldsymbol{W}_{\setminus i\setminus i})^T_{XZ}(\boldsymbol{\beta_i})_X = \boldsymbol{\beta_i}^T\boldsymbol{W}_{\setminus i\setminus i}^{ODB}\boldsymbol{\beta_i},
\end{equation*}
and considering that
\begin{align*}
    & (\boldsymbol{\beta_i})_X^T(\boldsymbol{W}_{\setminus i\setminus i})_{XX}(\boldsymbol{\beta_i})_X + (\boldsymbol{\beta_i})_Z^T(\boldsymbol{W}_{\setminus i\setminus i})_{ZZ}(\boldsymbol{\beta_i})_Z \\
    & = \boldsymbol{\beta_i}^T\boldsymbol{W}_{\setminus i\setminus i}\boldsymbol{\beta_i} - (\boldsymbol{\beta_i})_X^T(\boldsymbol{W}_{\setminus i\setminus i})_{XZ}(\boldsymbol{\beta_i})_Z - (\boldsymbol{\beta_i})_Z^T(\boldsymbol{W}_{\setminus i\setminus i})^T_{XZ}(\boldsymbol{\beta_i})_X \\
    & = \boldsymbol{\beta_i}^T\boldsymbol{W}_{\setminus i\setminus i}\boldsymbol{\beta_i} - \boldsymbol{\beta_i}^T\boldsymbol{W}_{\setminus i\setminus i}^{ODB}\boldsymbol{\beta_i},
\end{align*}

the objective function becomes:
\begin{equation*}
    F_{\boldsymbol{\Lambda}, \boldsymbol{W}_{\setminus i\setminus i}}(\boldsymbol{\beta_i}) = \frac{1}{2}\boldsymbol{S}_i^T\boldsymbol{\Theta}_{\setminus i\setminus i}\boldsymbol{S}_i + \frac{1 + c}{2}\boldsymbol{\beta_i}^T\boldsymbol{W}_{\setminus i\setminus i}\boldsymbol{\beta_i} - c\boldsymbol{\beta_i}^T\boldsymbol{W}_{\setminus i\setminus i}^{ODB}\boldsymbol{\beta_i} - \boldsymbol{\beta_i}^T\boldsymbol{S}_i + \left \| \boldsymbol{\Lambda}_i \odot \boldsymbol{\beta}_i \right \|_1
\end{equation*}

Now we minimize with respect to the $j$th element of $\boldsymbol{\beta_i}$, symbolized here by $\beta_j$. Of the dot products in the previous equation, we display in the next only the terms dependent on $\beta_j$. Moreover, we remark that all the penalty coefficients that are part of $\boldsymbol{\Lambda}$ are defined to be positive.
\begin{align*}
    \frac{\partial F_{\boldsymbol{\Lambda}, \boldsymbol{W}_{\setminus i\setminus i}}(\boldsymbol{\beta_i})}{ \partial \beta_j} & = \frac{\partial}{\partial \beta_j} \left \{ \frac{1 + c}{2}\left(\beta_j^2(\boldsymbol{W}_{\setminus i\setminus i})_{jj}+2\beta_j \sum_{k \neq j}(\boldsymbol{W}_{\setminus i\setminus i})_{jk}\beta_k\right) \right . \\
    & - c\left(\beta_j^2(\boldsymbol{W}_{\setminus i\setminus i})_{jj}^{ODB} +2\beta_j \left . \sum_{k \neq j}(\boldsymbol{W}_{\setminus i\setminus i})_{jk}^{ODB}\beta_k\right) - \beta_jS_{ji} + \left \| \boldsymbol{\Lambda}_i \odot \boldsymbol{\beta}_i \right \|_1 \right \} \\
    & = (1+c)\beta_j(\boldsymbol{W}_{\setminus i\setminus i})_{jj} + (1+c)\sum_{k \neq j}(\boldsymbol{W}_{\setminus i\setminus i})_{jk}\beta_k - 2c\sum_{k \neq j}(\boldsymbol{W}_{\setminus i\setminus i})_{jk}^{ODB}\beta_k \\
    & - S_{ji} + \Lambda_{ji}\frac{\partial |\beta_j|}{\partial \beta_j}
\end{align*}

Notice how the term containing $(\boldsymbol{W}_{\setminus i\setminus i})_{jj}^{ODB}$ disappears, as all diagonal elements of $\boldsymbol{W}_{\setminus i\setminus i}^{ODB}$ are zero. In particular, with $j$ being part of the generic dataset $A$, that could either be $X$ or $Z$:
\begin{align*}
    (1+c)\sum_{k \neq j}(\boldsymbol{W}_{\setminus i\setminus i})_{jk}\beta_k & = (1+c)\sum_{k \notin A}(\boldsymbol{W}_{\setminus i\setminus i})_{jk}\beta_k + (1+c)\sum_{ \substack{k \in A \\ k \neq j}}(\boldsymbol{W}_{\setminus i\setminus i})_{jk}\beta_k \\
    2c\sum_{k \neq j}(\boldsymbol{W}_{\setminus i\setminus i})_{jk}^{ODB}\beta_k & = 2c\sum_{k \notin A}(\boldsymbol{W}_{\setminus i\setminus i})_{jk}^{ODB}\beta_k = 2c\sum_{k \notin A}(\boldsymbol{W}_{\setminus i\setminus i})_{jk}\beta_k
\end{align*}

Again, here all the elements $(\boldsymbol{W}_{\setminus i\setminus i})_{jk}^{ODB}$ are zero for every $k \in A$, since the diagonal blocks of the matrix are zero, and all the other elements coincide with the corresponding elements in $\boldsymbol{W}_{\setminus i\setminus i}$. Hence, setting  $\frac{\partial F_{\boldsymbol{\Lambda}, \boldsymbol{W}_{\setminus i\setminus i}}(\boldsymbol{\beta_i})}{ \partial \beta_j} = 0$, we can rewrite the equation as:
\begin{align*}
    \beta_j & = \frac{1}{(1+c)(\boldsymbol{W}_{\setminus i\setminus i})_{jj}}\left\{S_{ji} - (1+c) \sum_{k \notin A}(\boldsymbol{W}_{\setminus i\setminus i})_{jk}\beta_k - (1+c)\sum_{ \substack{k \in A \\ k \neq j}}(\boldsymbol{W}_{\setminus i\setminus i})_{jk}\beta_k \right.\\
    & \left. +2c \sum_{k \notin A}(\boldsymbol{W}_{\setminus i\setminus i})_{jk}\beta_k - \Lambda_{ji}\frac{\partial |\beta_j|}{\partial \beta_j}\right\},
\end{align*}

which, for $\alpha=\frac{1}{1+c}$ and simplifying, can be written as: 
\begin{align*}
    \beta_j & = \frac{1}{(\boldsymbol{W}_{\setminus i\setminus i})_{jj}}\left\{ \alpha S_{ji} - \sum_{ \substack{k \in A \\ k \neq j}}(\boldsymbol{W}_{\setminus i\setminus i})_{jk}\beta_k  - \alpha(1-c) \sum_{k \notin A}(\boldsymbol{W}_{\setminus i\setminus i})_{jk}\beta_k - \alpha\Lambda_{ji}\frac{\partial |\beta_j|}{\partial \beta_j}\right\},
\end{align*}

From here, by applying the soft-threshold operator as seen in \citet{friedman_pathwise_2007}, one can straightforwardly obtain Algorithm 1 as shown in Section 3.

\section*{Appendix B: Equivalences involving the estimated variance-covariance matrix}

It can be proven that the squared root of a symmetric positive semidefinite matrix is a symmetric positive semidefinite matrix itself. Nevertheless, such proof is out of the scope of this manuscript. Given this statement, since both $\boldsymbol{W}_{\setminus i\setminus i}$ and $\boldsymbol{\Theta}_{\setminus i\setminus i}$ are symmetric positive semidefinite, so are their squared roots. This allows us to prove the following two sets of equivalences:

\begin{align}
    & \begin{pmatrix}
        (\boldsymbol{W}_{\setminus i\setminus i})_{XX} & (\boldsymbol{W}_{\setminus i\setminus i})_{XZ}\\
(\boldsymbol{W}_{\setminus i\setminus i})^T_{XZ} & (\boldsymbol{W}_{\setminus i\setminus i})_{ZZ}
    \end{pmatrix} = \boldsymbol{W}_{\setminus i\setminus i} = \boldsymbol{W}_{\setminus i\setminus i}^{1/2}\boldsymbol{W}_{\setminus i\setminus i}^{1/2} = (\boldsymbol{W}_{\setminus i\setminus i}^{1/2})^T\boldsymbol{W}_{\setminus i\setminus i}^{1/2} \nonumber \\
    & = \begin{pmatrix}
        (\boldsymbol{W}_{\setminus i\setminus i}^{1/2})_X^T \\
        (\boldsymbol{W}_{\setminus i\setminus i}^{1/2})_Z^T
    \end{pmatrix} \begin{pmatrix}
        (\boldsymbol{W}_{\setminus i\setminus i}^{1/2})_X &
        (\boldsymbol{W}_{\setminus i\setminus i}^{1/2})_Z
    \end{pmatrix} = \begin{pmatrix}
        (\boldsymbol{W}_{\setminus i\setminus i}^{1/2})_X^T(\boldsymbol{W}_{\setminus i\setminus i}^{1/2})_X & (\boldsymbol{W}_{\setminus i\setminus i}^{1/2})_X^T(\boldsymbol{W}_{\setminus i\setminus i}^{1/2})_Z\\
        (\boldsymbol{W}_{\setminus i\setminus i}^{1/2})_Z^T(\boldsymbol{W}_{\setminus i\setminus i}^{1/2})_X & (\boldsymbol{W}_{\setminus i\setminus i}^{1/2})_Z^T(\boldsymbol{W}_{\setminus i\setminus i}^{1/2})_Z
    \end{pmatrix}
\end{align}

\begin{equation}
    \Theta_{\setminus i\setminus i} = \boldsymbol{W}_{\setminus i\setminus i}^{-1}=\boldsymbol{W}_{\setminus i\setminus i}^{-1/2}\boldsymbol{W}_{\setminus i\setminus i}^{-1/2} = (\boldsymbol{W}_{\setminus i\setminus i}^{-1/2})^T\boldsymbol{W}_{\setminus i\setminus i}^{-1/2}
\end{equation}

\clearpage

\begin{figure}[!p]
\centering\includegraphics[width=0.8\textwidth]{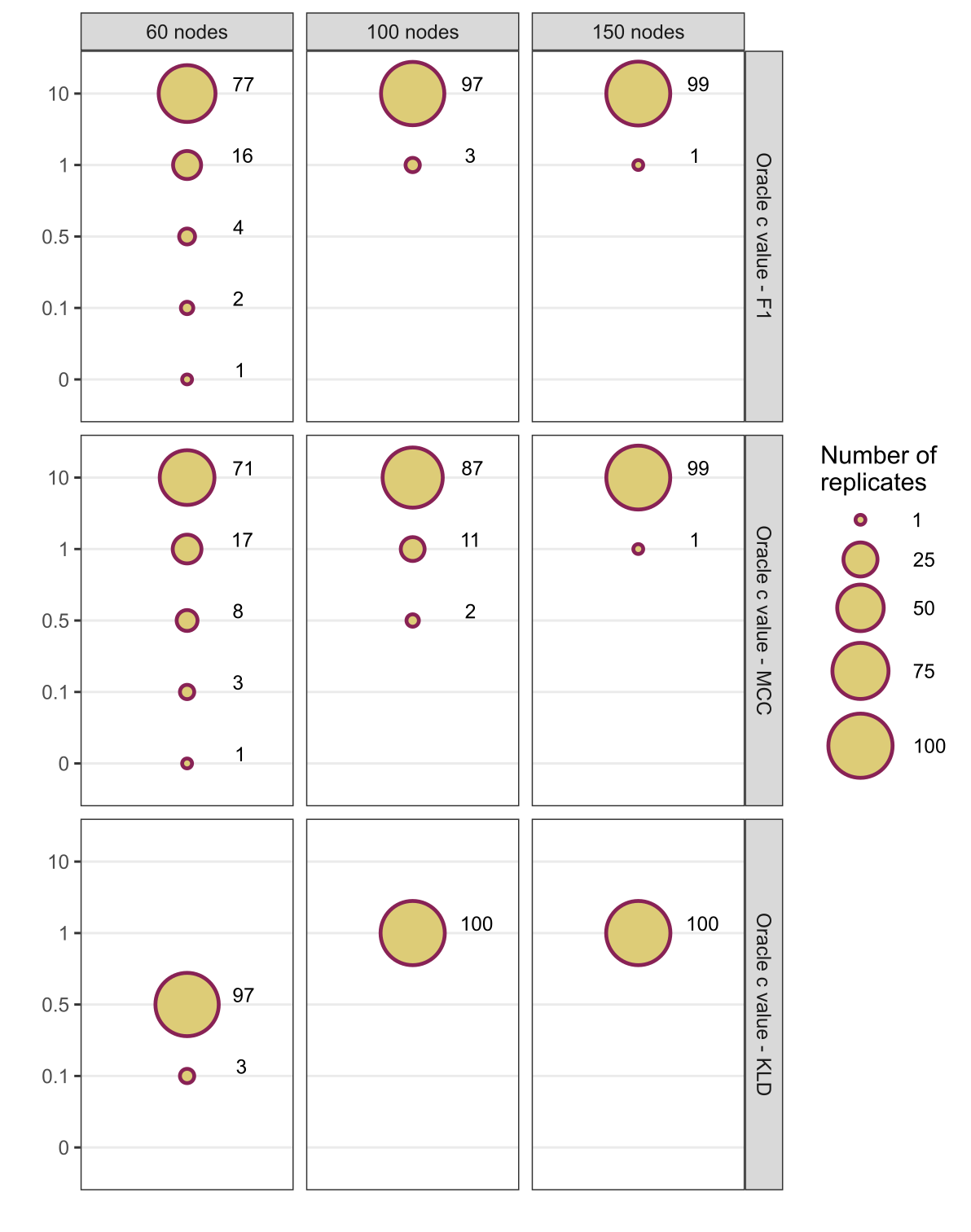}
\caption{Bubble plot of the $c$ value of oracle \textit{coglasso} networks in our simulated scenarios and according to the three measures used. The simulation scenarios are distributed along the columns, and the measures are distributed along rows. The size of the bubbles represents the proportion of the $100$ replicates of each simulation scenario for which the $c$ value indicated on the y-axis led to the best achieving \textit{coglasso} network according to the corresponding performance measure. For each bubble, we report the associated number of replicates beside it. The $c$ value is zero only for a single replicate and only according to $F_1$ and \emph{MCC} in the least complex scenario. In the $99$ remaining replicates of the least complex scenario for $F_1$ and \emph{MCC}, and in all replicates for \emph{KLD}, the $c$ value of the best network is never zero. In both the more complex scenarios, the $c$ of the best network is also never zero. This implies a role of collaboration in achieving the best network.}
\label{fig:oracle_c}
\end{figure}

\begin{figure}[!p]
\centering\includegraphics[width=1.1\textwidth]{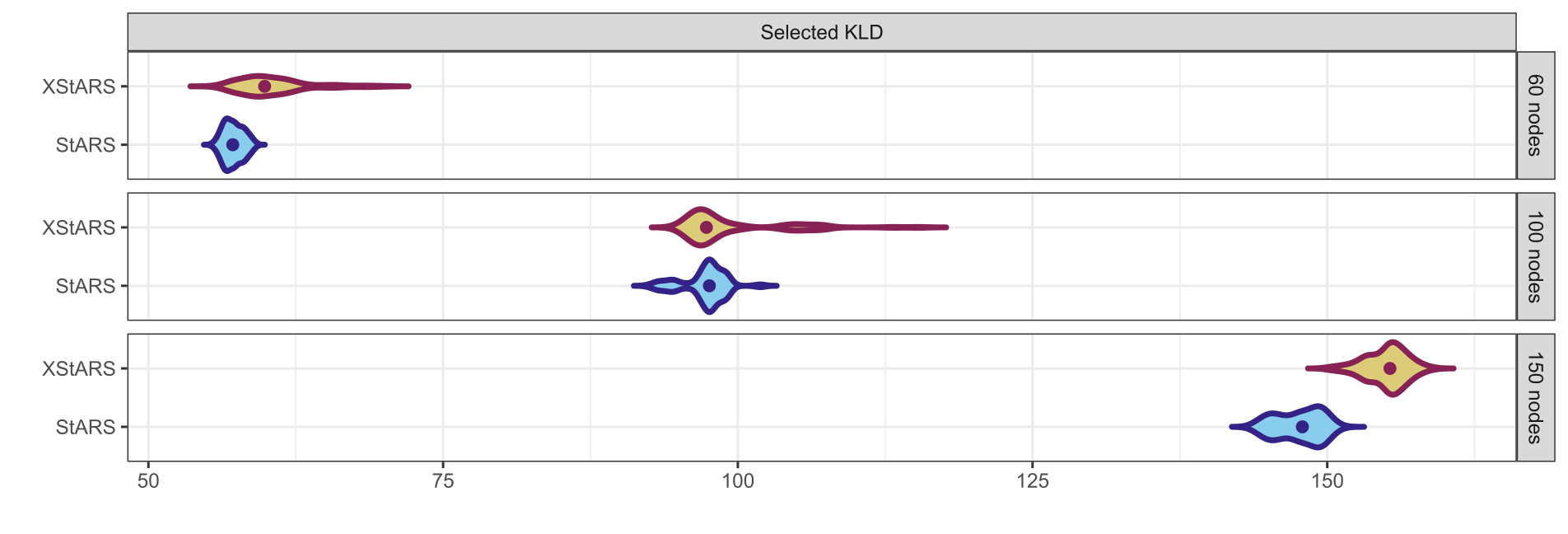}
\caption{Performance of model selection with \textit{XStARS} for \textit{coglasso} (in yellow-red) and \textit{StARS} for \textit{glasso} (in cyan-blue) over three increasingly complex and high-dimensional scenarios (networks with $60$, $100$, and $150$ nodes) in terms of Kullback-Leibler divergence (\emph{KLD}). The scenarios are distributed along the rows of the figure. The violins of the figure are composed of $100$ data points, one for each replicate of the simulations, and the networks were reconstructed from datasets of $n = 50$ observations.}
\label{fig:selected_KLD}
\end{figure}

\begin{figure}[!p]
\centering\includegraphics[width=0.8\textwidth]{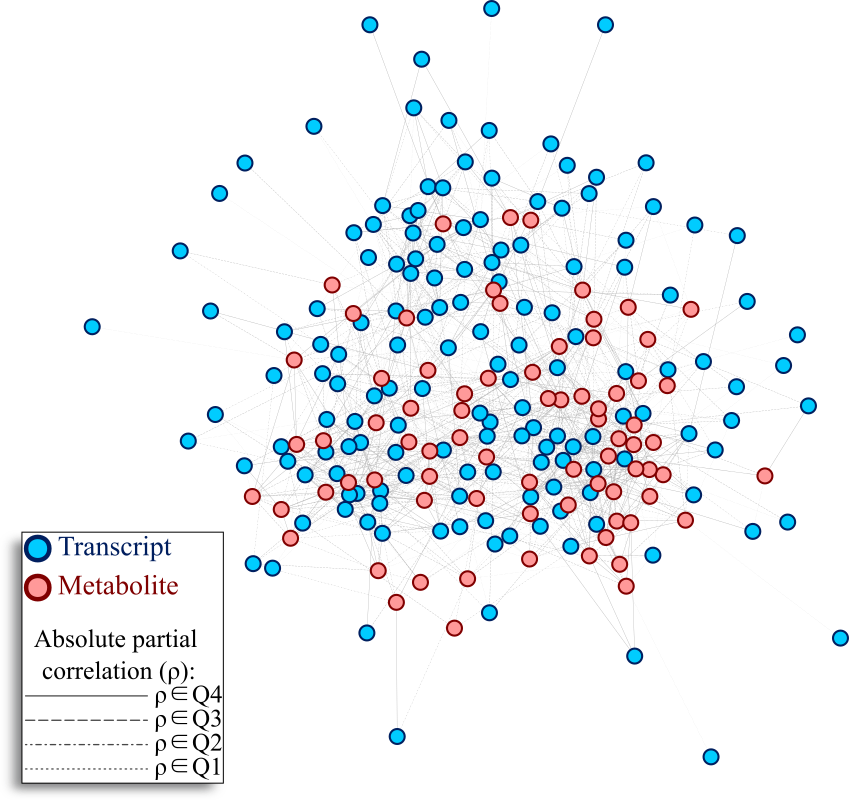}
\caption{Multi-omics network inferred using \textit{coglasso}, selected by \textit{XStARS} model selection. Blue nodes represent transcripts, while pink nodes represent metabolites. The fragmentation of an edge represents the absolute partial correlation related to the edge, according to the quartile they belong to. The strongest partial correlations are part of the fourth quartile, and are represented as continuous lines. The lower quartiles are progressively represented with more fragmented lines, until the fully dotted line of the first quartile. The figure shows how the two omics layers are highly blended, representing a large amount of connections shared among them.}
\label{fig:network}
\end{figure}

\begin{figure}[!p]
\centering\includegraphics[width=0.8\textwidth]{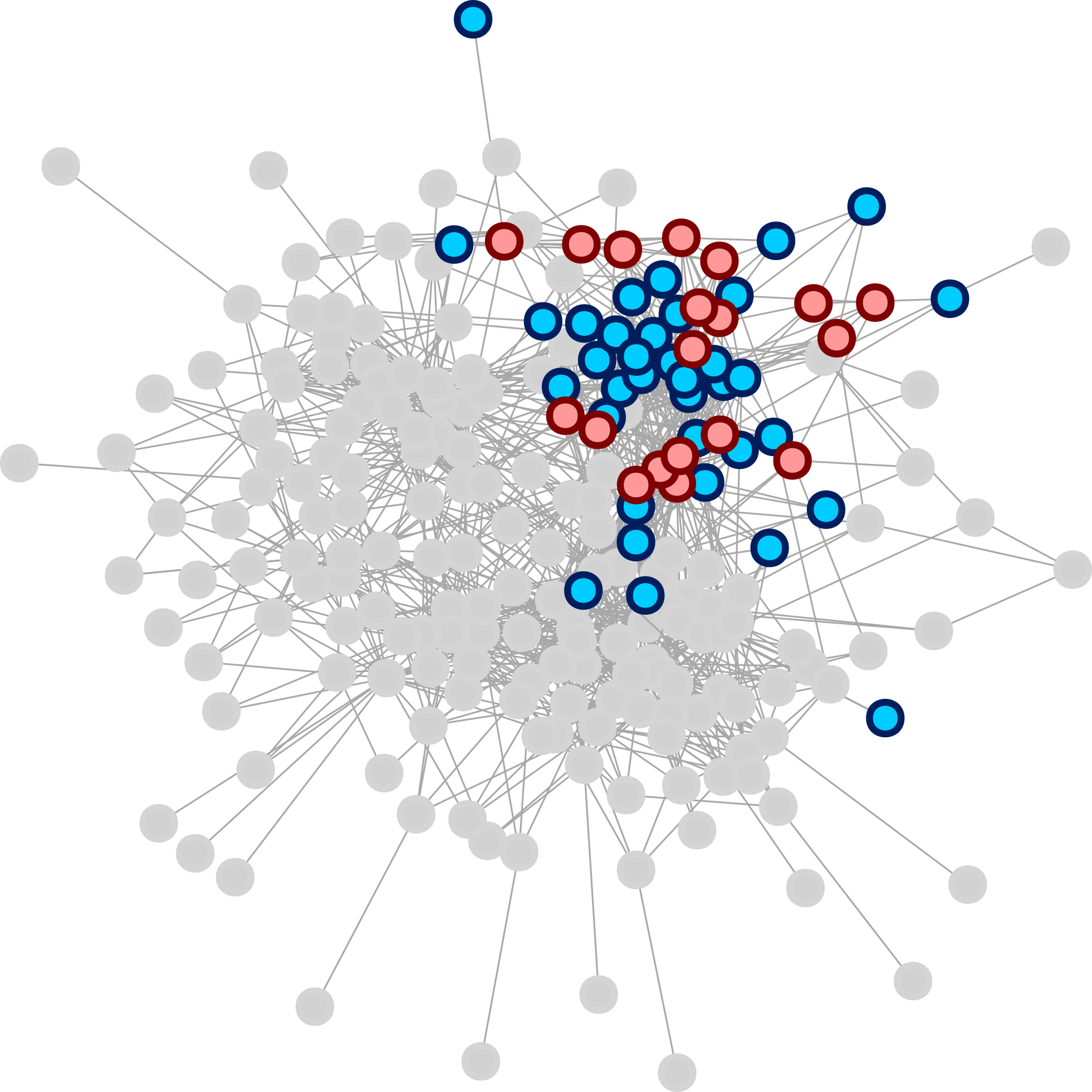}
\caption{Multi-omics network inferred using \textit{coglasso}. The nodes belonging to the second largest node community are highlighted here.}
\label{fig:comm2}
\end{figure}

\bibliographystyle{plainnat}
\bibliography{coglasso}